\documentclass[final,3p,times,twocolumn,longtitle]{elsarticle}

\usepackage{amssymb}
\usepackage{color}
\usepackage{geometry}
\usepackage{epstopdf}

\usepackage{lineno}

\journal{Astroparticle Physics}

\begin{document}

\begin{frontmatter}



\title{Energy Calibration of CALET Onboard the International Space Station} 


\author[AF6,AF5]{Y.~Asaoka\corref{cor1}}

\author[AF2,AF3]{Y.~Akaike}
\author[AF0]{Y.~Komiya}
\author[AF0]{R.~Miyata}
\author[AF6,AF5,AF0]{S.~Torii}
\author[AF1,AF40,AF41]{O.~Adriani}
\author[AF4]{K.~Asano}
\author[AF7,AF40]{M.G.~Bagliesi}
\author[AF7,AF40]{G.~Bigongiari}
\author[AF8]{W.R.~Binns}
\author[AF7,AF40]{S.~Bonechi}
\author[AF1,AF40,AF41]{M.~Bongi}
\author[AF7,AF40]{P.~Brogi}
\author[AF8]{J.H.~Buckley}
\author[AF9]{N.~Cannady}
\author[AF1,AF40,AF41]{G.~Castellini}
\author[AF10]{C.~Checchia}
\author[AF9]{M.L.~Cherry}
\author[AF10]{G.~Collazuol}
\author[AF11,AF40]{V.~Di~Felice}
\author[AF12]{K.~Ebisawa}
\author[AF12]{H.~Fuke}
\author[AF9]{T.G.~Guzik}
\author[AF13,AF3]{T.~Hams}
\author[AF14]{M.~Hareyama}
\author[AF6]{N.~Hasebe}
\author[AF15]{K.~Hibino}
\author[AF16]{M.~Ichimura}
\author[AF17]{K.~Ioka}
\author[AF4]{W.~Ishizaki}
\author[AF8]{M.H.~Israel}
\author[AF9]{A.~Javaid}
\author[AF6]{K.~Kasahara}
\author[AF6]{J.~Kataoka}
\author[AF18]{R.~Kataoka}
\author[AF19]{Y.~Katayose}
\author[AF20]{C.~Kato}
\author[AF21]{N.~Kawanaka}
\author[AF22]{Y.~Kawakubo}
\author[AF23]{H.~Kitamura}
\author[AF8]{H.S.~Krawczynski}
\author[AF2,AF3]{J.F.~Krizmanic}
\author[AF16]{S.~Kuramata}
\author[AF24]{T.~Lomtadze}
\author[AF7,AF40]{P.~Maestro}
\author[AF7,AF40]{P.S.~Marrocchesi}
\author[AF24]{A.M.~Messineo}
\author[AF25]{J.W.~Mitchell}
\author[AF26]{S.~Miyake}
\author[AF27]{K.~Mizutani}
\author[AF38,AF3]{A.A.~Moiseev}
\author[AF6,AF12]{K.~Mori}
\author[AF28]{M.~Mori}
\author[AF1,AF40,AF41]{N.~Mori}
\author[AF39]{H.M.~Motz}
\author[AF20]{K.~Munakata}
\author[AF6]{H.~Murakami}
\author[AF12]{Y.E.~Nakagawa}
\author[AF5]{S.~Nakahira}
\author[AF12]{J.~Nishimura}
\author[AF15]{S.~Okuno}
\author[AF29]{J.F.~Ormes}
\author[AF6]{S.~Ozawa}
\author[AF1,AF40,AF41]{L.~Pacini}
\author[AF11,AF40]{F.~Palma}
\author[AF1,AF40,AF41]{P.~Papini}
\author[AF7,AF37]{A.V.~Penacchioni}
\author[AF8]{B.F.~Rauch}
\author[AF1,AF40,AF41]{S.~Ricciarini}
\author[AF38,AF3]{K.~Sakai}
\author[AF22]{T.~Sakamoto}
\author[AF13,AF3]{M.~Sasaki}
\author[AF15]{Y.~Shimizu}
\author[AF30]{A.~Shiomi}
\author[AF11,AF40]{R.~Sparvoli}
\author[AF1,AF40,AF41]{P.~Spillantini}
\author[AF7,AF40]{F.~Stolzi}
\author[AF31]{I.~Takahashi}
\author[AF12]{M.~Takayanagi}
\author[AF4]{M.~Takita}
\author[AF15]{T.~Tamura}
\author[AF15]{N.~Tateyama}
\author[AF32]{T.~Terasawa}
\author[AF12]{H.~Tomida}
\author[AF33]{Y.~Tsunesada}
\author[AF23]{Y.~Uchihori}
\author[AF12]{S.~Ueno}
\author[AF1,AF40,AF41]{E.~Vannuccini}
\author[AF9]{J.P.~Wefel}
\author[AF34]{K.~Yamaoka}
\author[AF35]{S.~Yanagita}
\author[AF22]{A.~Yoshida}
\author[AF36]{K.~Yoshida}
\author[AF4]{T.~Yuda}

\address[AF6]{Research Institute for Science and Engineering, Waseda University, 3-4-1 Okubo, Shinjuku, Tokyo 169-8555, Japan}
\address[AF5]{JEM Mission Operations and Integration Center, Human Spaceflight Technology Directorate, 
Japan Aerospace Exploration Agency, 2-1-1 Sengen, Tsukuba, Ibaraki 305-8505, Japan}
\address[AF2]{Universities Space Research Association, 7178 Columbia Gateway Drive Columbia, MD 21046, USA.}
\address[AF3]{CRESST and Astroparticle Physics Laboratory NASA/GSFC, Greenbelt, MD 20771, USA}
\address[AF0]{School of Advanced Science and Engineering, Waseda University, 3-4-1 Okubo, Shinjuku, Tokyo 169-8555, Japan}
\address[AF1]{University of Florence, Via Sansone, 1 - 50019 Sesto, Fiorentino, Italy}
\address[AF4]{Institute for Cosmic Ray Research, The University of Tokyo, 5-1-5 Kashiwa-no-Ha, Kashiwa, Chiba 277-8582, Japan}
\address[AF7]{University of Siena, Rettorato, via Banchi di Sotto 55, 53100 Siena, Italy}
\address[AF8]{Department of Physics,Washington University, One Brookings Drive, St. Louis, MO 63130-4899, USA}
\address[AF9]{Department of Physics and Astronomy, Louisiana State University, 202 Nicholson Hall, Baton Rouge, LA 70803, USA}
\address[AF10]{Department of Physics and Astronomy, University of Padova, Via Marzolo, 8, 35131 Padova, Italy}
\address[AF11]{University of Rome Tor Vergata, Via della Ricerca Scientifica 1, 00133 Rome, Italy}
\address[AF12]{Institute of Space and Astronautical Science, Japan Aerospace Exploration Agency, 
3-1-1 Yoshinodai, Chuo, Sagamihara, Kanagawa 252-5210, Japan}
\address[AF13]{Department of Physics, University of Maryland, Baltimore County, 1000 Hilltop Circle, Baltimore, MD 21250, USA}
\address[AF14]{St. Marianna University School of Medicine, 2-16-1, Sugao, Miyamae-ku, Kawasaki, Kanagawa 216-8511, Japan}
\address[AF15]{Kanagawa University, 3-27-1 Rokkakubashi, Kanagawa, Yokohama, Kanagawa 221-8686, Japan}
\address[AF16]{Faculty of Science and Technology, Graduate School of Science and Technology, Hirosaki University, 
3, Bunkyo, Hirosaki, Aomori 036-8561, Japan}
\address[AF17]{Yukawa Institute for Theoretical Physics, Kyoto University, Kitashirakawa Oiwakecho, Sakyo, Kyoto 606-8502, Japan}
\address[AF18]{National Institute of Polar Research, 10-3, Midori-cho, Tachikawa, Tokyo 190-8518, Japan}
\address[AF19]{Faculty of Engineering, Division of Intelligent Systems Engineering, Yokohama National University, 
79-5 Tokiwadai, Hodogaya, Yokohama 240-8501, Japan}
\address[AF20]{Faculty of Science, Shinshu University, 3-1-1 Asahi, Matsumoto, Nagano 390-8621, Japan}
\address[AF21]{School of Science, The University of Tokyo, 7-3-1 Hongo, Bunkyo, Tokyo 113-003, Japan}
\address[AF22]{College of Science and Engineering, Department of Physics and Mathematics, Aoyama Gakuin University,  
5-10-1 Fuchinobe, Chuo, Sagamihara, Kanagawa 252-5258, Japan}
\address[AF23]{National Institute of Radiological Sciences, 4-9-1 Anagawa, Inage, Chiba 263-8555, Japan}
\address[AF24]{University of Pisa and INFN, Italy}
\address[AF25]{Astroparticle Physics Laboratory, NASA/GSFC, Greenbelt, MD 20771, USA}
\address[AF26]{Department of Electrical and Electronic Systems Engineering, National Institute of Technology, Ibaraki College, 866 Nakane, 
Hitachinaka, Ibaraki 312-8508 Japan}
\address[AF27]{Saitama University, Shimo-Okubo 255, Sakura, Saitama, 338-8570, Japan}
\address[AF28]{Department of Physical Sciences, College of Science and Engineering, Ritsumeikan University, Shiga 525-8577, Japan}
\address[AF29]{Department of Physics and Astronomy, University of Denver, Physics Building, Room 211, 2112 East Wesley Ave., Denver, CO 
80208-6900, USA}
\address[AF30]{College of Industrial Technology, Nihon University, 1-2-1 Izumi, Narashino, Chiba 275-8575, Japan}
\address[AF31]{Kavli Institute for the Physics and Mathematics of the Universe, The University of Tokyo, 5-1-5 Kashiwanoha, Kashiwa, 277-8583, Japan}
\address[AF32]{RIKEN, 2-1 Hirosawa, Wako, Saitama 351-0198, Japan}
\address[AF33]{Division of Mathematics and Physics, Graduate School of Science, Osaka City University, 3-3-138 Sugimoto, Sumiyoshi, 
Osaka 558-8585, Japan}
\address[AF34]{Nagoya University, Furo, Chikusa, Nagoya 464-8601, Japan}
\address[AF35]{Graduate School of Science and Engineering, Ibaraki University, 2-1-1 Bunkyo, Mito, Ibaraki 310-8512, Japan}
\address[AF36]{Department of Electronic Information Systems, Shibaura Institute of Technology, 307 Fukasaku, Minuma, Saitama 337-8570, Japan}
\address[AF37]{ASI Science Data Center (ASDC), Via del Politecnico snc, 00133 Rome, Italy}
\address[AF38]{Department of Astronomy, University of Maryland, College Park, Maryland 20742, USA }
\address[AF39]{International Center for Science and Engineering Programs, Waseda University, 3-4-1 Okubo, Shinjuku, Tokyo 169-8555, Japan}
\address[AF40]{National Institute for Nuclear Physics (INFN), Piazza dei Caprettari, 70 - 00186 Rome, Italy}
\address[AF41]{Institute of Applied Physics (IFAC),  National Research Council (CNR), Via Madonna del Piano, 10, 50019 Sesto, Fiorentino, Italy}
\cortext[cor1]{Corresponding author. \\ E-mail address: yoichi.asaoka@aoni.waseda.jp (Y.Asaoka)}

\begin{abstract}
In August 2015, the CALorimetric Electron Telescope (CALET), 
designed for long exposure observations of high energy cosmic rays, 
docked with the International Space Station (ISS) and shortly thereafter began to
collect data. 
CALET will measure the cosmic ray electron spectrum over the energy range of 1~GeV to 20~TeV 
with a very high resolution of 2\% above 100~GeV, based on 
a dedicated instrument incorporating an exceptionally thick 30 radiation-length 
calorimeter with both total absorption and imaging (TASC and IMC) units. 
Each TASC readout channel must be carefully 
calibrated over the extremely wide dynamic range of CALET that spans six orders of magnitude
in order to obtain a degree of calibration accuracy matching the resolution of energy measurements. 
These calibrations consist of calculating the conversion factors between ADC units and energy deposits, 
ensuring linearity over each gain range, 
and providing a seamless transition between neighboring gain ranges. 
This paper describes these calibration methods in detail, along with the resulting data  and associated accuracies. %
The results presented in this paper 
show that a sufficient accuracy was achieved for the calibrations of each channel in order to obtain a suitable
resolution over the entire dynamic range of the electron spectrum measurement. 

\end{abstract}


\end{frontmatter}

\section{Introduction}
\label{intro}
\begin{figure}[hbt!]
\begin{center}
\includegraphics[width=0.95\hsize]{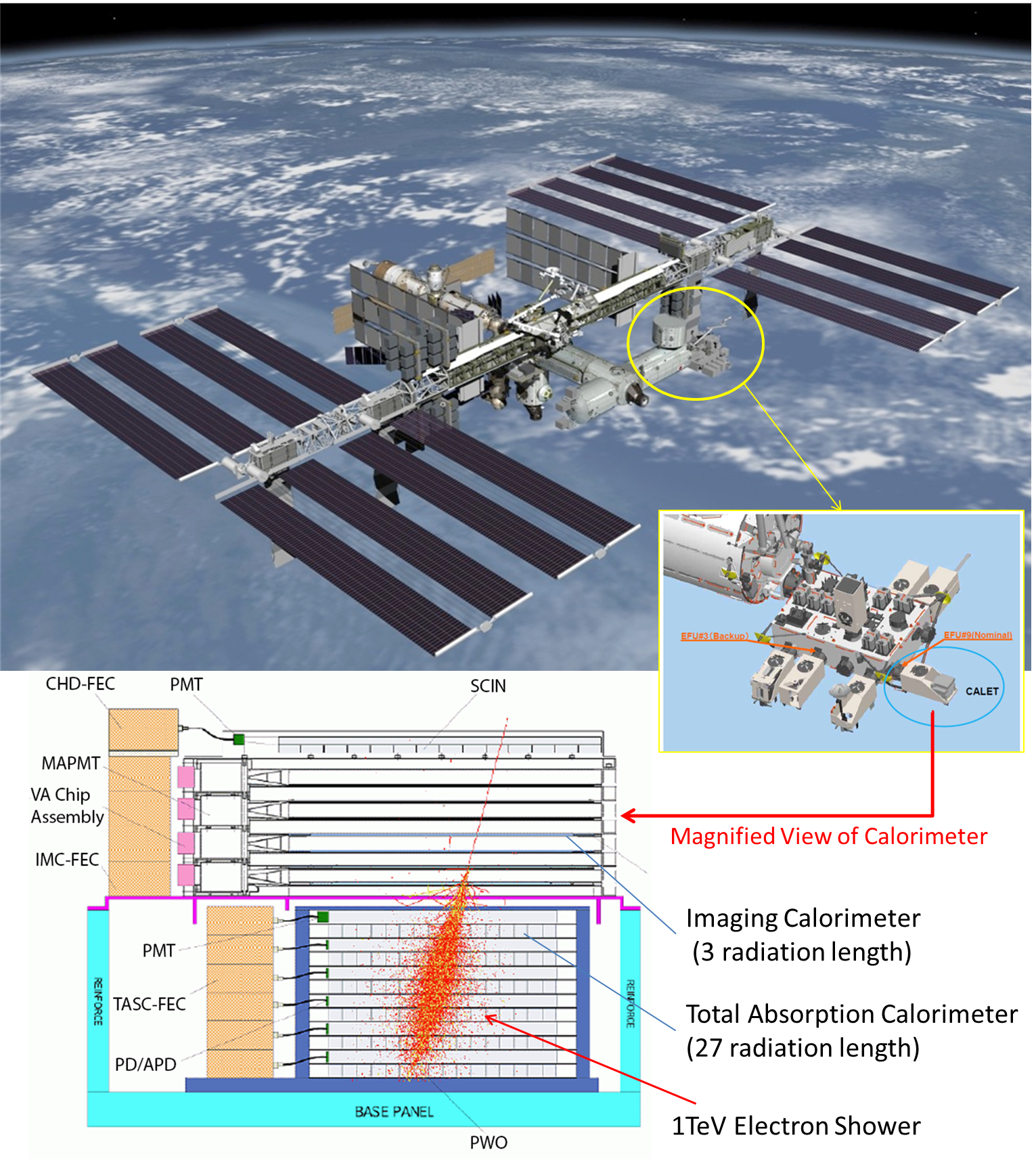}
\caption{The CALET detector onboard the ISS as part of the Japanese Experiment Module - Exposed Facility~\cite{caletICRC2015}.}
\label{fig:calet}.

\end{center}
\end{figure}

The CALET (CALorimetric Electron Telescope)~\cite{caletICRC2015}
was docked to 
Exposed Facility of the Japanese Experiment Module (JEM-EF)
on the International Space Station (ISS) in August 2015 
and
has been collecting data~\cite{wcocICRC2015} since October 2015. 
It has been designed for long duration 
observations of high energy cosmic rays onboard the ISS. 
The CALET detector, shown in Fig.~\ref{fig:calet}, includes a 
very thick calorimeter unit of 30 radiation-length ($X_{0}$), 
consisting of imaging and total absorption
calorimeters (IMC and TASC, respectively). 
The primary purpose of CALET is to make full use 
of a total-containment 
and well-segmented calorimeter to discover 
nearby cosmic-ray accelerators~\cite{Kobayashi2004,Kawanaka2011} 
and to search for dark matter~\cite{Motz2015} 
with precision measurements of electron and 
gamma ray spectra over a wide energy range. 

The calorimeter absorbs the majority of an electron shower's energy in the TeV energy 
range and is able to identify electrons within a very high proton flux, with a rejection  
factor of $>10^5$, based on the difference in shower development. 
This instrument will therefore be used to acquire the cosmic ray electron spectrum over the energy range of 
1 GeV to 20 TeV with exceptional energy resolution, especially above 100~GeV, where the resolution is better than 2\%. 
Since each channel of the lead tungstate (PbWO$_{4}$) crystals of the TASC 
has a dynamic range of six order of magnitudes, CALET is capable 
of determining the energy of primary particles from 1 GeV to 1 PeV.
This enables the instrument to measure proton and nuclei spectra as well as electron and 
gamma ray spectra over this extremely wide energy range. 

The cosmic-ray detectors based on magnetic spectrometers that are presently in use (PAMELA \cite{Pamela-e} and AMS-02 \cite{AMS02-e}) 
have the significant advantage of being able to distinguish the sign of the charge on the particle. 
However, the spectral observations of these devices are
limited to energy values below $\sim$1~TeV because their detector systems are optimized for the observation of 
various cosmic rays that have energies below this value.
In addition, previous calorimeter-type instruments (ATIC \cite{ATIC-e} and Fermi-LAT \cite{Fermi-e})
were not optimized for the observation of electrons, and so at present their ability to identify electrons  
in the presence of a very high proton flux at higher energies is limited. In contrast, 
CALET is fully capable of measuring the electron plus positron spectrum well into the TeV region, 
as the result of being equipped with a thick 30~$X_0$ calorimeter.
Due to its extremely high energy 
resolution and its ability to discriminate electrons from hadrons, 
CALET will allow the detailed search for 
various spectral structures of high-energy electron cosmic rays, 
perhaps providing the first
experimental evidence of the presence of a nearby astrophysical cosmic-ray source. 
Even though it cannot distinguish the charge sign, 
CALET has the potential to detect distinctive features in the TeV region of electron plus positron energy spectrum 
possibly resulting from dark matter annihilation/decay. 
In the opposite scenario, 
the information acquired by CALET should make it possible to set significantly more stringent limits 
on dark matter annihilation compared to current experimental data \cite{Motz2015}. 

There is an intrinsic advantage in measuring the electromagnetic components of cosmic rays with CALET. 
Since the TASC absorbs the majority of the energy ($\sim$95\%) contained in an  
electromagnetic cascade, well into the TeV region, CALET is able to measure 
the primary energies of cosmic ray electrons and gamma rays with very small corrections.
In principle, this should result in precise energy measurements with very
low systematic errors.
However, in order to achieve a calibration accuracy that matches the intrinsic energy resolution over
the wide dynamic range of six orders of magnitude, 
a careful calibration of each 
TASC readout channel is required. 
The present paper details the calibration methods and results 
and also summarizes the accuracy of resulting energy measurements. 

This paper is organized as follows.
In Section~\ref{inst}, the CALET instrumentation is briefly summarized, 
while the energy measurements and calibration methods are described in Section~\ref{method}.
Section~\ref{calib} presents each step of the CALET energy calibration process 
in detail, along with the resulting data. 
In Section~\ref{disc}, the calibration accuracy is studied and its effects 
on the energy resolution and absolute scale are assessed. 
Last, a summary and conclusions are presented in Section~\ref{conc}. 

\section{CALET Instrumentation}
\label{inst}
The CALET calorimeter, shown in side view in the lower part of Fig.~\ref{fig:calet} along with a simulation of a 1 TeV electron shower, has several unique and important characteristics, as briefly noted in the Introduction. 
These include its ability to resolve in detail 
the initial development of showers, as well as 
tracks generated by non-interacting minimum ionizing particles (MIPs), 
and its capacity to precisely measure the energy of electrons in the TeV region as a result of 
its depth of 30~$X_0$. 
These features are achieved through a combination of three primary
detector sub-systems: Particle identification and energy measurements are performed by the TASC, charge identification is obtained from a charge detector (CHD), and an imaging calorimeter (IMC) is employed for track reconstruction.
The key performance of each detector component, 
as described below, was estimated on the basis of a detailed Monte Carlo (MC) simulation and was confirmed by several beam tests carried out primarily at 
the CERN-SPS facility.

Plastic scintillators arranged in two orthogonal layers, each containing 14 scintillator paddles (3.2 $\times$ 1.0 $\times$ 44.8 cm$^3$), 
constitute the CHD. 
These paddles generate photons that are detected by a 
photomultiplier tube (PMT), and the resulting output is sent to a front
end circuit (FEC). 
This FEC and the subsequent readout system have sufficient dynamic range 
for particle charges in the range of $Z = 1 \sim 40$. 
The charge resolution of the CHD spans the range from 0.15 electron charge units ($e$) for boron to $\simeq 0.30-0.35 e$ in the iron region \cite{Pier13}. 

The initial shower is visualized by the IMC sampling calorimeter, 
which has been carefully designed to accurately determine the shower starting point and incident direction. 
This calorimeter has a thickness of 3 $X_{0}$ and contains five upper 0.2 $X_{0}$ and two lower 1.0 $X_{0}$ tungsten plates. 
The IMC contains a total of 16 detection layers, arranged in 8 X-Y pairs, with each layer segmented into
448 parallel scintillating fibers (0.1 $\times$ 0.1 $\times$ 44.8 cm$^3$), which are read out by
64-channel multi anode PMTs. 
By reconstructing the trajectory of incident particles in the IMC, 
the arrival direction of each individual particle can be determined.
Above several tens of GeV, the expected angular resolution for gamma-rays 
is $\sim 0.24^{\circ}$, 
while the angular resolution for electrons is better and close to $\sim 0.16^{\circ}$ \cite{caletICRC2015}. 

The TASC has an overall depth of 27 $X_{0}$ and consists of 12 detection layers
in an alternating orthogonal arrangement, 
each comprised of 16 lead tungstate crystal (PbWO$_4$ or PWO) logs 
with dimensions of 2.0 $\times$ 1.9 $\times$ 32.6 cm$^3$. 
As a result of this design, the TASC is able to image the development of the shower in three dimensions. 
With the exception of the first layer which uses PMTs, 
a photodiode (PD) in conjunction with an avalanche photodiode (APD) reads the photons generated by each PWO log. 
Employing dual shaping amplifiers with two different gains for each APD (PMT) and PD, 
increases the dynamical range to 10$^6$ (10$^4$).
As a consequence, the TASC can measure the energy of the incident electrons and gamma rays 
with a resolution $<2$\% above 100 GeV \cite{Niita15}. 
Another important role of the TASC is to efficiently identify high-energy electrons among the overwhelming background of cosmic-ray protons. 
Particle identification information from both the IMC and TASC is used to achieve an electron detection efficiency above 80 \% and a proton rejection power of $\sim 10^{5}$ \cite{Akaike2011}.

A preselected combination of 
simultaneous trigger counter signals, 
which are produced by discriminating the analog signals 
from the detector components, 
generates an event-trigger decision. 
As such, 
each of CHD X and Y, IMC X1--X4, Y1--Y4 and TASC X1 generates lower discriminator
signals \cite{wcocICRC2015}. 
The signals from two IMC fiber layers are processed by a single front end circuit, 
and so each axis has only four trigger counter signals. 
Three trigger modes are possible in CALET. 
The High Energy (Low Energy) Trigger 
select shower events with energies greater than 10~GeV (1~GeV), 
while the Single Trigger is 
dedicated to acquiring data from non-interacting particles for the purposes of detector calibration.
\section{Energy Measurement and Calibration Method}
\label{method}
As briefly introduced in Section~\ref{intro}, 
careful calibration of each
TASC readout channel is required 
in order to achieve a calibration accuracy that matches the intrinsic energy resolution over
the wide dynamic range of six orders of magnitude.
The entire dynamic range is covered by four different
gain ranges, based on two photon detectors -  
an APD and a PD - 
in conjunction with a shaper amplifier with lower and higher gains. 
The energy calibration process consists of three steps as follows:
\begin{enumerate}
\item determination of the conversion factor between ADC units and the energy deposit, 
\item linearity measurements over each gain range, 
\item correlation measurements between adjacent gain ranges.
\end{enumerate}

The first step is the calibration of the energy deposit of each channel to obtain an ADC unit-to-energy conversion factor using MIPs, 
known as MIP calibration. 
As it is the case with other detectors intended for direct cosmic-ray measurements, 
CALET can use penetrating particles to equalize the gains of different detector segments, 
based on the fact that the
energy deposits of such particles in the relativistic energy range 
are approximately constant.
In contrast with the calibration of a spectrometer, 
MIP calibration serves as an absolute energy 
calibration of the CALET because this instrument is a total absorption calorimeter.
Therefore, absolute end-to-end energy calibrations are possible using 
the MIP technique. End-to-end calibration stands for the summary 
treatment of all detector responses transforming the particle's energy 
loss to the output signal, such as PWO scintillation yield, photon 
propagation in the PWO, quantum efficiency of the APD/PD, gain of the 
front end circuit and others.

Prior to launch, 
the linearity over each gain range was confirmed by on-ground calibration 
using a UV pulse laser, during which the APD and PD outputs were determined as a function of 
the laser energy. In this process, the UV laser pulse was absorbed by the PWO, while the APD and PD detected 
the resulting scintillation emissions resulting from de-excitation of the PWO. 
By scanning the UV laser pulses over 
the entire energy range, it was therefore possible to calibrate the input-to-output 
correspondence for all four gain ranges. 

The adjacent gain ranges were subsequently cross-calibrated based on their gain ratios.
Taking advantage of the nearly one order of magnitude overlap between adjacent
gain ranges, it was possible to measure identical energy deposits within two
gain ranges.
In contrast to the MIP calibration process, which requires a dedicated trigger mode, 
for gain correlation measurements all the data regardless of trigger modes can be used. 
This is helpful especially for PD range correlation measurements because such measurements require
a long term accumulation of data for very high energy events. 
As the linearity of each gain range had already been confirmed, 
the gain over entire dynamic range could be determined based on 
the ADC-to-energy conversion factor, using the acquired gain ratios between adjacent gain ranges. 
This process automatically takes into account possible gain changes due to the flight environment. 
Such gain changes are expected to occur due to variations in temperature between flight and ground calibration. Special care was also required to account separately for the effects of 
the flight environment on the APD gain and the light yield of the PWO, which in turn affects both the APD and PD gain ranges. 

\section{Energy Calibration}
\label{calib}
\subsection{MIP Calibration}
\label{mip}
It is an important advantage of the CALET instrument that an absolute end-to-end 
calibration of the energy scale is possible, by employing the MIP technique.
While 10\% accuracy is relatively easy to achieve using MIPs, more  
accurate calibration requires careful analysis of the 
energy distribution of incident particles, 
appropriate penetrating particle event selection, and
consideration of the position and temperature dependence of each TASC log.
The latter is especially important because CALET employs a one-end readout system 
and because of the relatively high temperature dependence of both the PWO and APD. This aspect of the calibration process is
discussed in detail in the following section.

While the energy deposits of relativistic particles
are approximately constant and close to minimum
ionization, the sample used for MIP calibration also
includes particles outside the minimum ionizing region.
Their energy spectrum depends on the cutoff rigidity,
and hence the geomagnetic latitude. As a result, the
position of the MIP peaks will shift by several percent
as a function of the geomagnetic latitude \cite{Niita15}.
To account for this effect, 
the incident particle energy distributions are assessed by simulating 
the energy spectra of incoming primary particles~\cite{Niita15} 
using ATMNC3~\cite{atmnc3},
in which AMS-01 proton and helium spectra \cite{AMS-01-pHe} were
used as input, since these data were taken at various geomagnetic 
latitudes, as well as them being in good agreement with BESS \cite{BESS98-pHe} and 
recent experiments.
As well, contamination by interacting particles and/or scattered and stopped particles can 
cause a systematic shift in the determined position of MIP peaks. 
In order to avoid this, 
the appropriate selection of penetrating particle events is 
ensured using a likelihood analysis~\cite{Niita15}.
To further improve the selection efficiency and to reduce systematic bias
during event selection, the likelihood ratios of
penetrating particles to interacting particles are also employed. 
By taking the ratios, the separation of 
penetrating particles from interacting particles becomes 
better, while possibly remaining discrepancies between flight
and MC data have less influence.

Event selection based on likelihood uses energy deposit distributions obtained from an MC simulation including the detector response of each channel. 
Simulating this detector response
requires data regarding the noise levels 
in units of energy, 
which in turn requires the ADC unit-to-energy conversion factor. 
Because this conversion factor is obtained from the calibration,
the MIP calibration is performed as an iterative procedure. 
However, this process converges very quickly and 
a single iteration is sufficient to obtain stable results when calibrating CALET.

\subsubsection{Position and Temperature Dependence Corrections}
\label{postemp}
\begin{figure}[hbt!]
\begin{center}
\includegraphics[width=0.99 \linewidth]{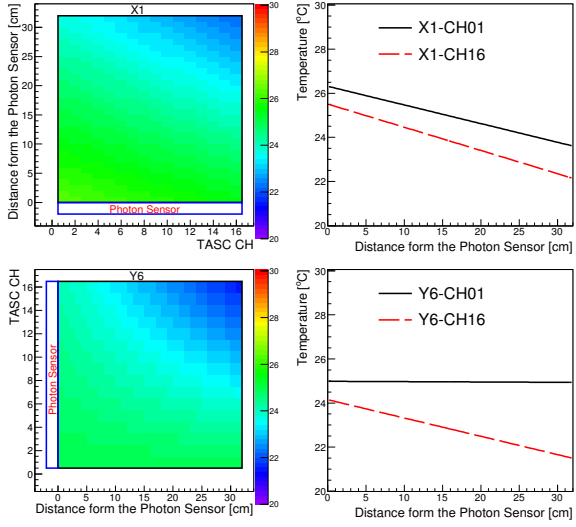}
\caption{
	Temperature distributions in the TASC X1 ({\it Top Left}) 
        and Y6 ({\it Bottom Left}) units, averaged over four months.
	Right side panels show the positional temperature variations along the length of each unit. 
	The black solid and red dashed lines in the {\it Top Right} ({\it Bottom Right}) panels 
	represent X1-CH01 and X1-CH16 (Y6-CH01 and Y6-CH16) data, 
	respectively.
	}
\label{fig:tasc_temp}
\end{center}
\end{figure}

To fully calibrate each TASC log, it is first necessary to 
correct for the position dependent effects so as to equalize the response along its length. 
In addition, because both the PWO light yield and the APD gain will vary with temperature, it is also required to correct for this temperature dependence. 
During the calibration process, the temperatures inside the TASC were calculated from 
temperature data measured during flight 
using a software that parameterized the temperature distribution in the TASC based on the CALET thermal model. 
The CALET flight model 
is equipped with 14 thermocouples located around the TASC structure. 
The CALET thermal model
was calibrated using the flight instrumentation results obtained from a 
thermal vacuum test performed at the JAXA Tsukuba Space Center.
Figure~\ref{fig:tasc_temp} presents the average temperature 
distributions inside the X1 ({\it Top}) and Y6 ({\it Bottom}) layers of the TASC.
Here, the left side panels show the two-dimensional temperature distributions, while 
the right side panels show the positional temperature dependence along the length of each unit.
Since it is not possible to 
differentiate between the gain change due to the general temperature 
slope and the inherent position dependence, the position dependence 
correction includes both effects.
These data clearly show that the temperature tends to decrease along the length of each unit.

\begin{figure}[tbh!]
\begin{center}
\includegraphics[width=0.95 \linewidth]{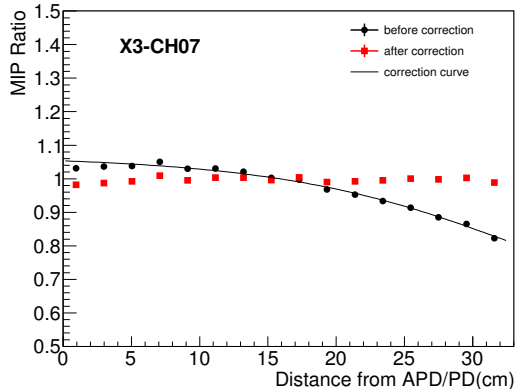}
\caption{
	An example of the position dependence of the MIP peak for a typical TASC log.
	The filled black and red circles represent data before and after 
	the correction, respectively. The black line indicates the function used to fit the position dependence.
	}
\label{fig:posdep}
\end{center}
\end{figure}

To correct for this position dependence, MIP peaks were determined for each of 16 segments defined along the length of each TASC log.
Subsequently, the position dependence of these MIP peaks for each log was fitted using an appropriate function of distance from the sensor (the PMT or APD/PD).
To ensure that the correct positional dependence was derived in each case, several different functions were defined.
Figure~\ref{fig:posdep} presents an example of the position dependence of a MIP peak
both before and after the correction process. 
On average, a position dependence of 9.2\% RMS was observed for a total of 192 PWO logs, 
and these data were successfully corrected. Following this correction, 
the RMS of the position dependence was reduced to 1.8\%. 

\begin{figure}[hbt!]
\begin{center}
\includegraphics[width=0.99 \linewidth]{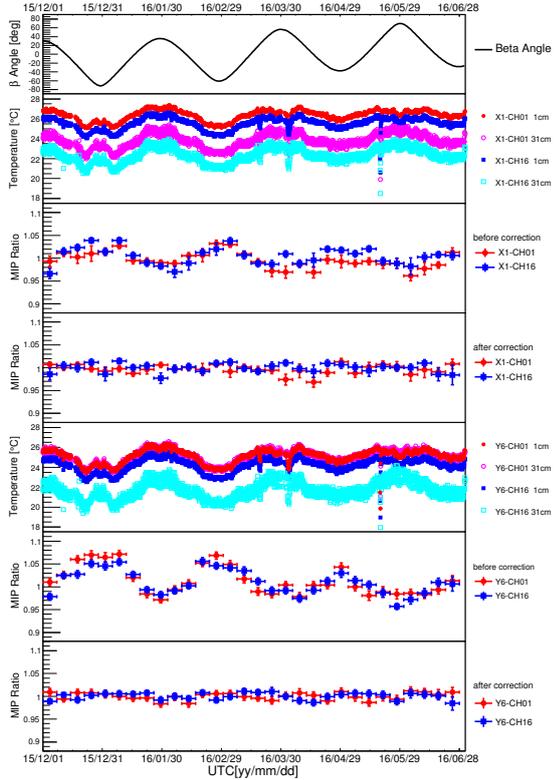}
\caption{The relationship between the beta angle, temperature and MIP peaks. 
The upper graph plots the time variation of the beta angle. 
The second graph shows the variations over time of the temperatures of four segments 
located in the corners of TASC X1. The third and fourth graphs represent 
the MIP variations of two different channels at both ends of TASC X1, 
before and after the temperature dependence correction. 
The last three graphs represent the same type of data for TASC Y6.} 
\label{fig:temp}
\end{center}
\end{figure}

In addition to the general temperature slopes in the TASC logs, there was also an overall 
temperature variation due to the dependence of temperature 
related to both the solar beta angle\footnote{solar beta angle is defined as the angle between 
the orbital plane of the ISS and the vector to the sun} 
and the solar altitude. 
To discriminate between these temperature variations and the position dependence 
due to temperature gradients in the TASC log data, 
we have obtained the averaged temperature at the center of each log and averaged
temperature gradient to calculate a position dependent reference
temperature for each track. 
The correction for temperature dependence then employed the difference from the reference temperature. 
In this manner, the data were corrected for both the beta angle dependence and the overall temperature changes due to solar 
altitude without interfering with the position dependence correction.
Figure ~\ref{fig:temp} presents examples of the overall temperature dependence of MIP peaks
over a period of seven months, together with solar beta angle variations over time (in the upper graph) and temperature variations 
for both the TASC X1 and Y6 layers.
These data indicate that the MIP peak variation rate due to the temperature changes was, on average, -1.9\% per degree for the PMT channels, and -3.4\% for the channels with an APD. 
Since these observed temperature dependences of the MIP peaks were 
consistent with one another within the associated errors, the average 
values for the PMT and APD were adopted as universal gain corrections 
independent of the PWO logs and reference temperatures.
Thanks to the performance of the active thermal control system (ATCS) available in the JEM-EF, 
temporal variations in the temperature were typically within a few degrees. 
On average, a temperature dependence of 3.3\% RMS was observed for 192 PWO logs, 
and this variation was successfully corrected for, reducing the  
RMS variation to 1.0\%. 

\subsubsection{Determination of the Energy Conversion Factor}
\begin{figure*}[htb!]
\begin{center}
\begin{minipage}{0.33\hsize}
\includegraphics[width=0.99 \linewidth]{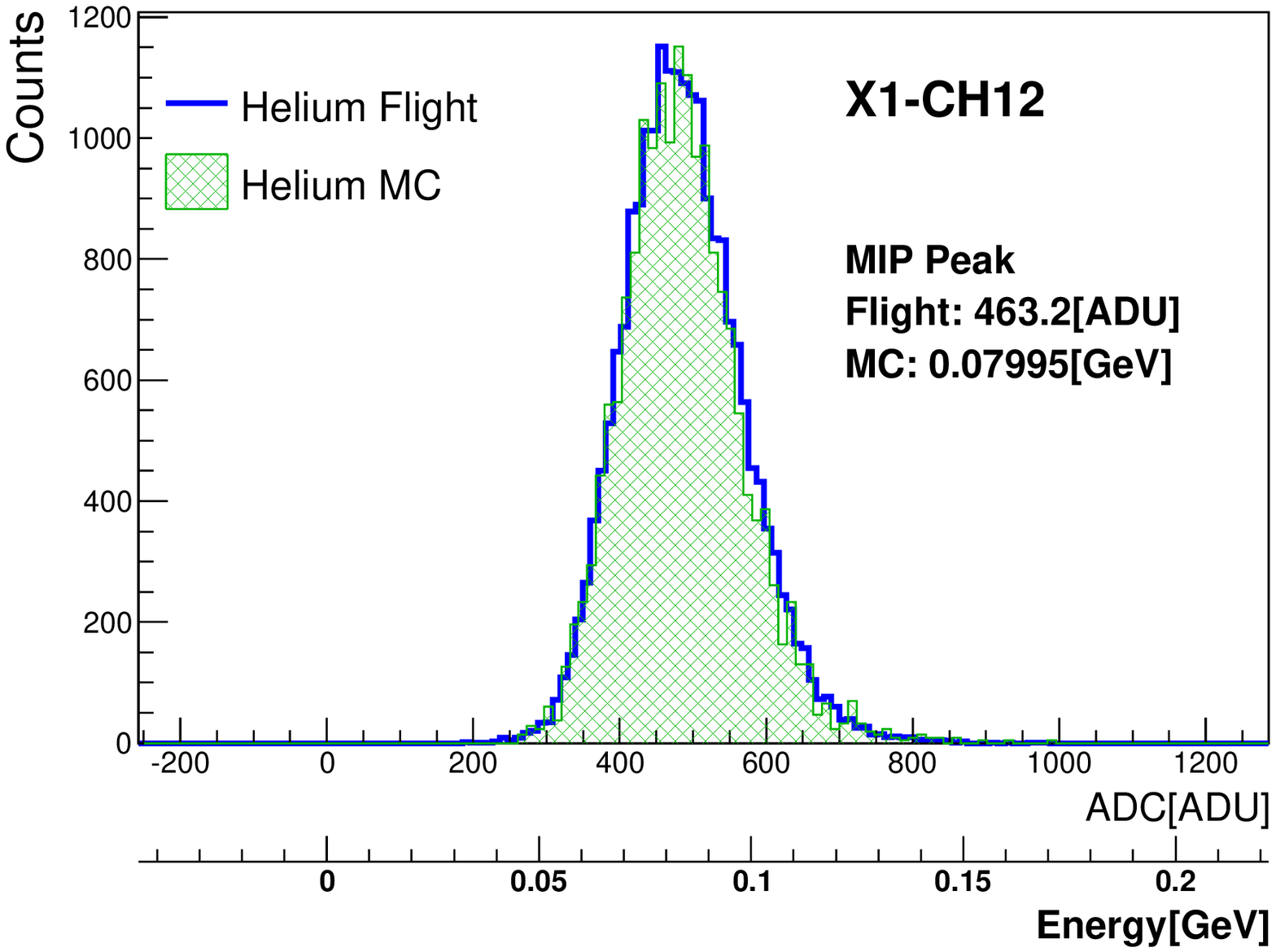} 
\end{minipage}
\begin{minipage}{0.33\hsize}
\includegraphics[width=0.99 \linewidth]{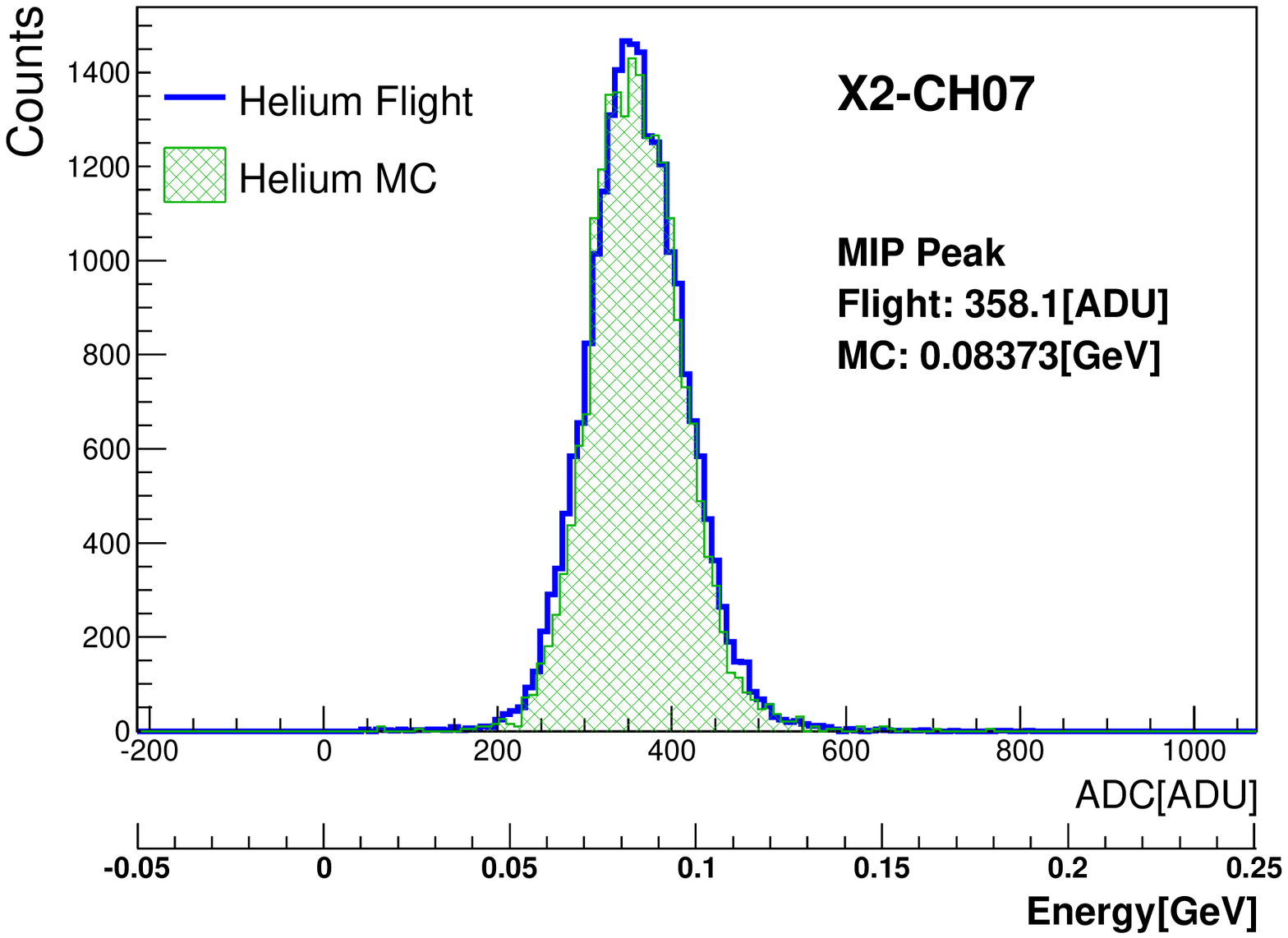} 
\end{minipage}
\begin{minipage}{0.33\hsize}
\includegraphics[width=0.99 \linewidth]{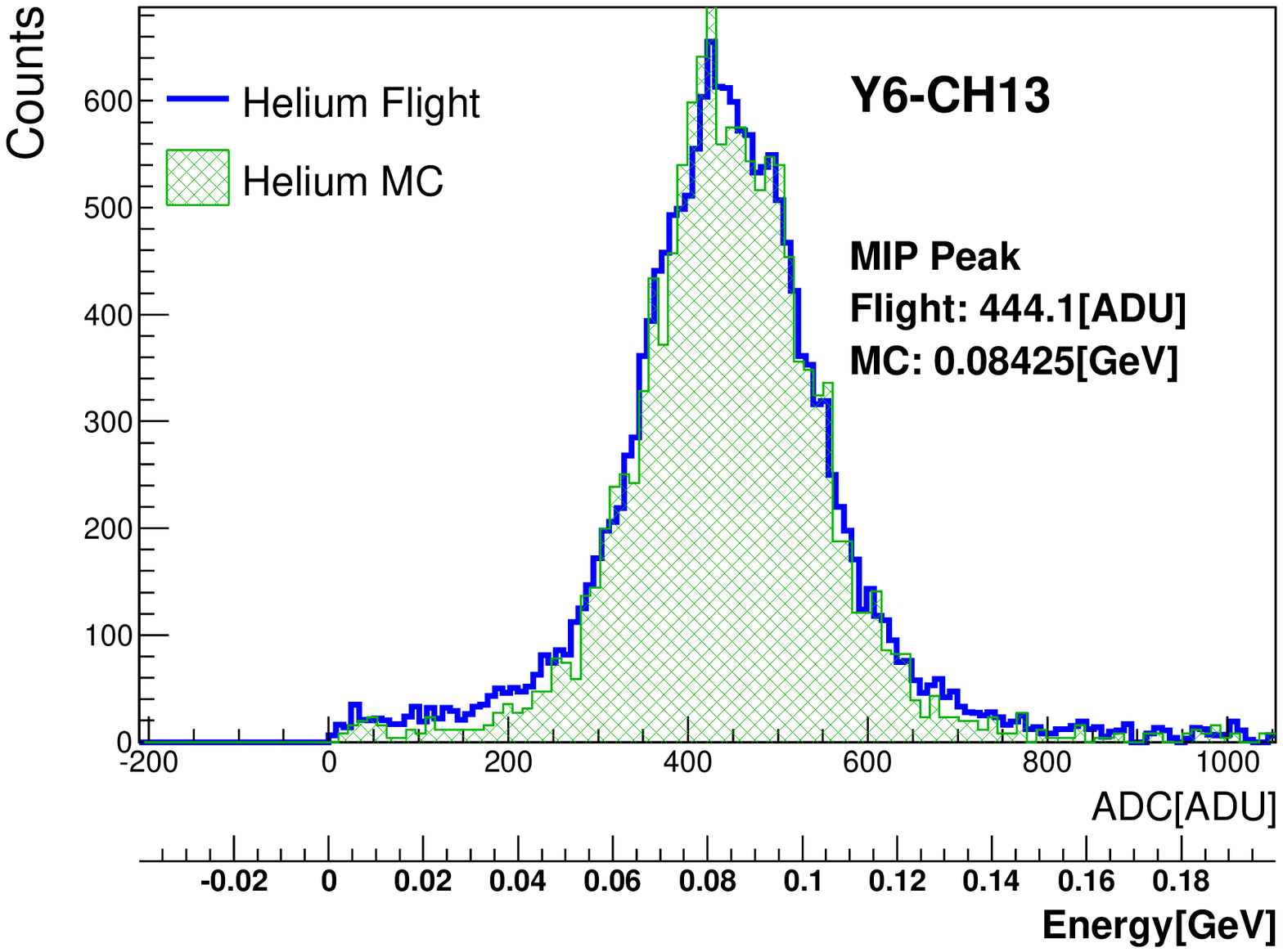} 
\end{minipage}
\begin{minipage}{0.33\hsize}
\includegraphics[width=0.99 \linewidth]{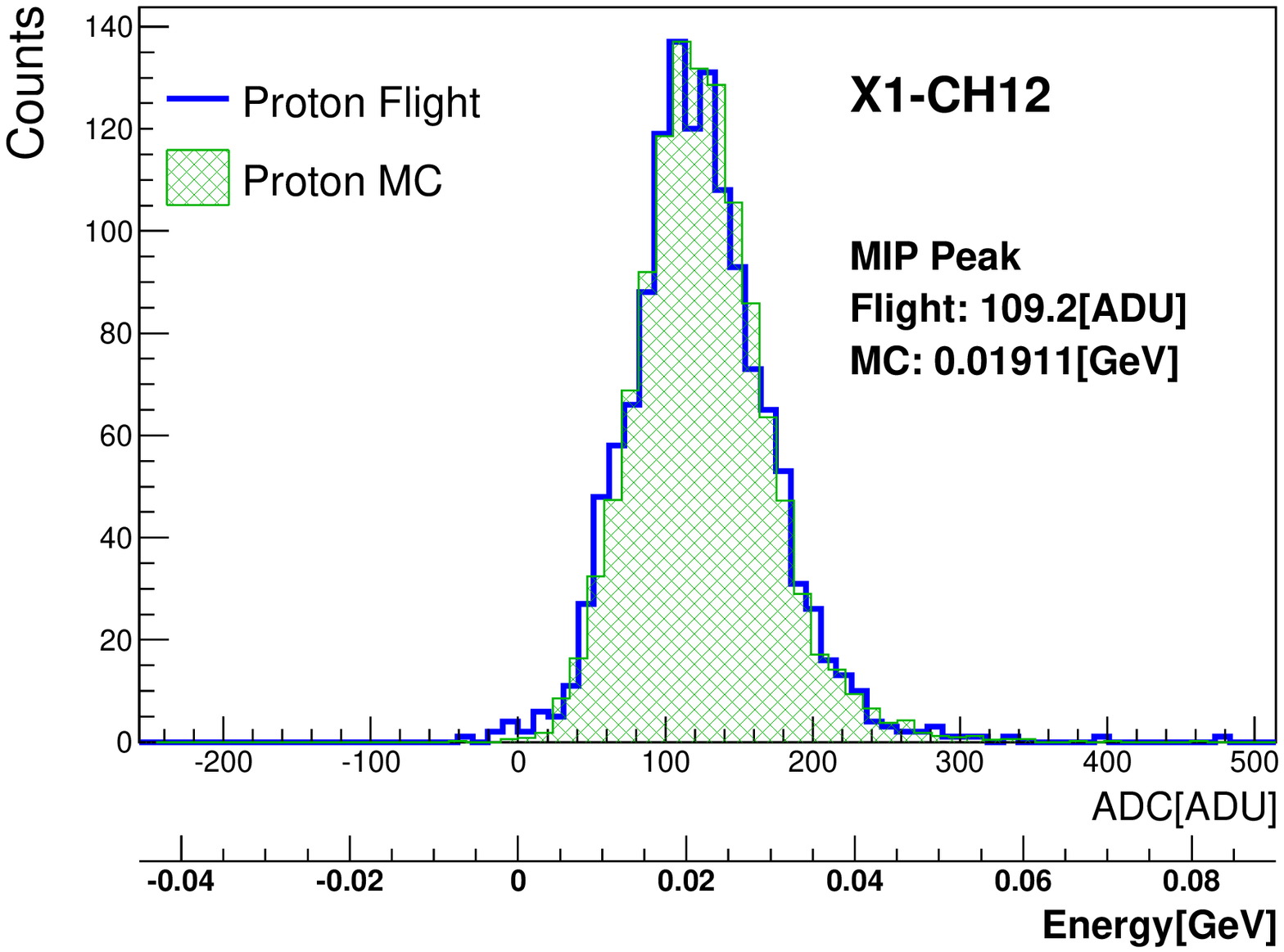} 
\end{minipage}
\begin{minipage}{0.33\hsize}
\includegraphics[width=0.99 \linewidth]{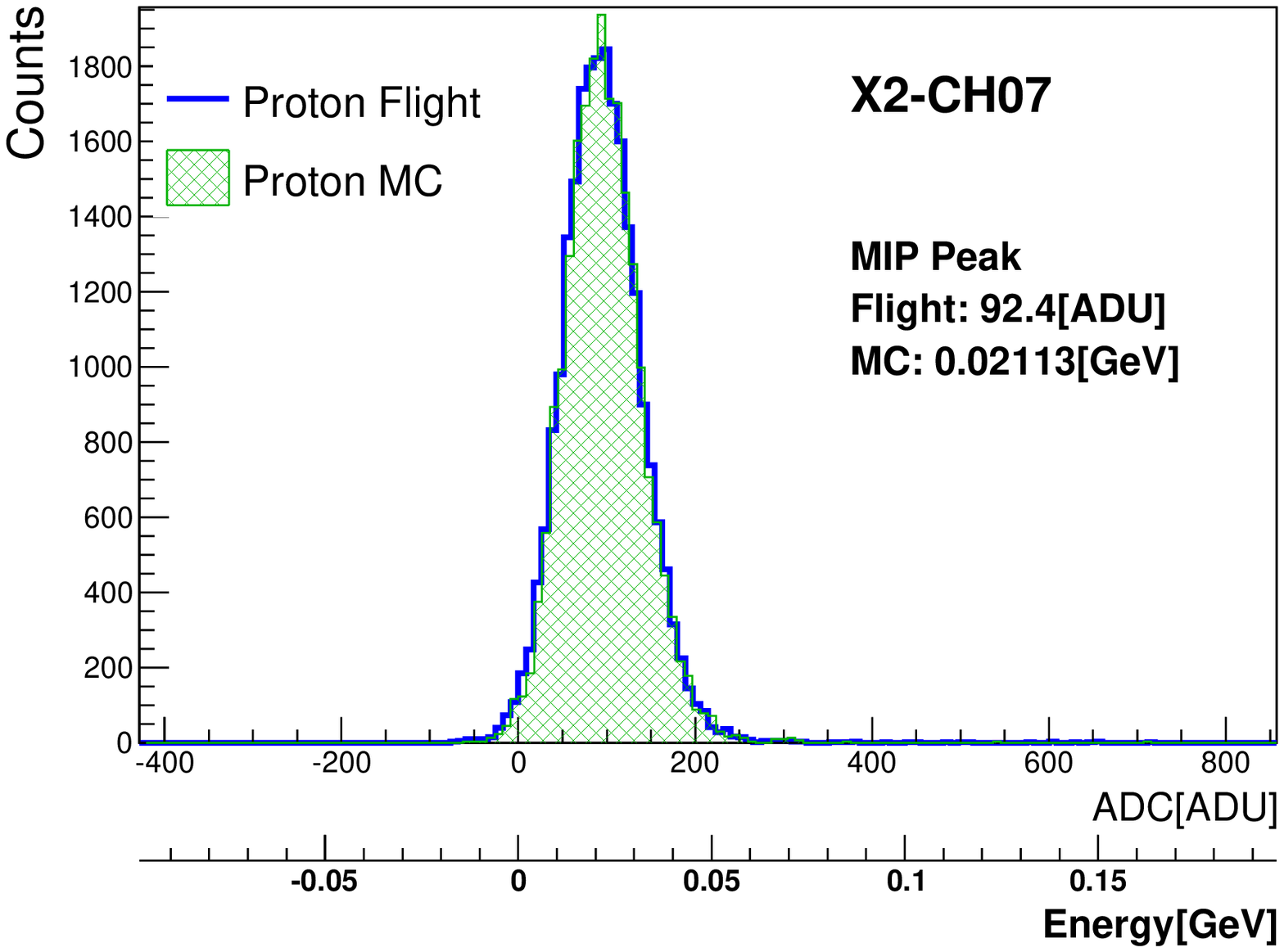} 
\end{minipage}
\begin{minipage}{0.33\hsize}
\includegraphics[width=0.99 \linewidth]{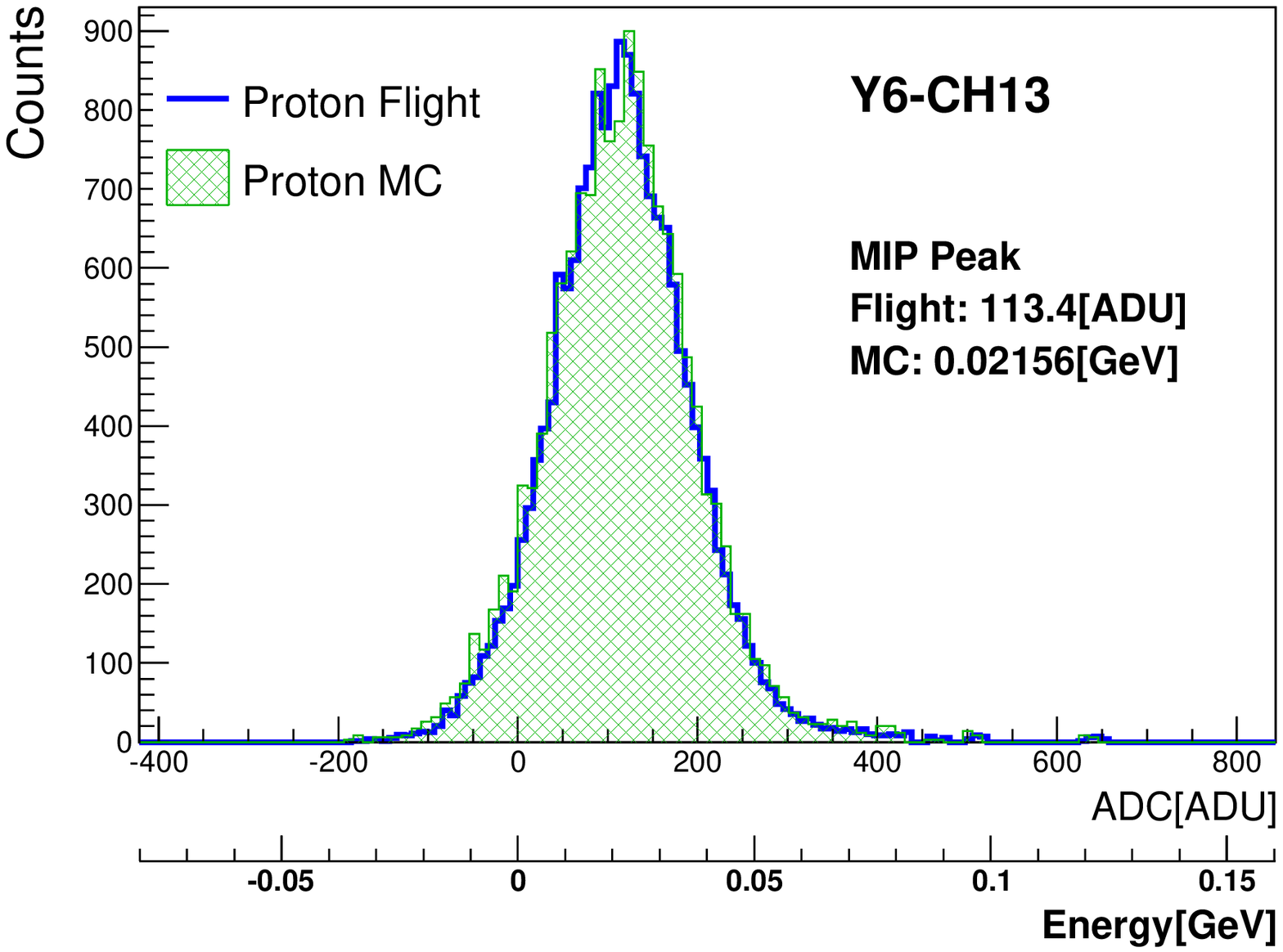} 
\end{minipage}
\caption{Comparisons of distributions of flight and simulated helium and proton data. 
Blue open and green hatched histograms represent flight and MC data, respectively. 
The top three plots provide helium distributions, while the bottom three show proton distributions. 
Data from one PMT channel ({\it Left}), one typical APD/PD channel ({\it Middle}) 
and one APD/PD channel in the bottom layer ({\it Right}) are shown.} 
\label{fig:mip}
\end{center}
\end{figure*}
Following the corrections for the position and temperature dependence described in 
Section~\ref{postemp}, accurate calculations of the MIP peaks in ADC units (ADU) could be 
obtained from the flight data, while MIP peaks in energy units could be 
determined from the simulated MC data. Subsequently, with the MIP peak values in both ADU and GeV, 
it was possible to find the energy conversion factor, GeV/ADU. In order to verify 
the accuracy of this conversion factor, factors were calculated 
for both proton and helium data.
As shown in Fig.~\ref{fig:mip}, clear peaks resulting from penetrating helium ({\it Top}) 
and protons ({\it Bottom}) were 
extracted using event selection based on likelihood analysis 
for both flight and MC event data. The MC event data were generated from 
a CALET detector simulation~\cite{Akaike2011} 
with the detector simulation tool EPICS~\cite{epics} 
using the ATMNC3 results as input data. 
The energy deposit of EPICS for PWO was
confirmed to be consistent within 1\% with the beam test data and with
the Geant4 results \cite{Karube11}.
In Fig.~\ref{fig:mip}, data from one PMT channel ({\it Left}), 
one typical APD/PD channel ({\it Middle}) and one APD/PD channel in the bottom layer ({\it Right}) are shown.
The conversion factor was calculated by comparing MIP distributions between 
flight and MC data; 
each distribution
was fitted with an appropriate function and the ratio of the peaks gave
the conversion factor. 
It is very important to properly smear the MC 
distribution according to the relevant noise factors. 
To do so, the Gaussian sigma of the fitted pedestal distribution of each TASC channel was used 
to incorporate electronic noise into the simulation. 
Fluctuations due to photoelectron statistics were included especially in the case of the TASC X1 channels equipped with PMTs, 
in addition to the pedestal noise, because such fluctuations have a significant effect 
due to the lower level of pedestal noise in these channels compared to 
APD channels.
The accuracy of each conversion factor was estimated from the errors in 
the peak fits on a channel-by-channel basis. 
On average, the accuracy values were 1.6\% and 0.6\% for protons and helium, respectively.
To ensure robustness of fit results, the fit range dependence of peak value was also investigated 
by changing the fit range by $\pm$33\% from its optimal value and it was found that such dependencies
were reasonably small as 0.4\% and 0.6\% for protons and helium, respectively. 
They are included in both calibration error and systematic uncertainty on the energy scale. 
Since the helium data have better statistics and a superior signal-to-noise ratio, it is evident that the more accurate determination 
of conversion factors was achieved using the helium data. 

Although this paper is focused on the calibration of the TASC,
the same method is applicable to the CHD and IMC, and in fact was employed 
when equalizing and calibrating the energy deposit of each of their channels.

\subsubsection{Estimation of Calibration Accuracy}
\begin{figure}[hbt!]
\begin{center}
\includegraphics[width=0.95 \linewidth]{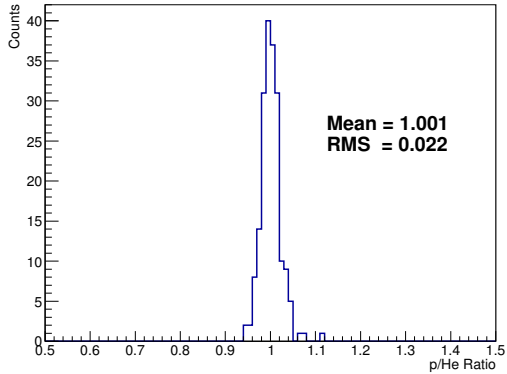}
\caption{Comparison of conversion factors obtained from proton and helium MIPs. 
The histogram represents the distribution of 
the proton MIP conversion factors divided by those obtained from helium MIPs
for each TASC log.}
\label{fig:phe}
\end{center}
\end{figure}

While it is relatively easy to estimate the accuracy of the calculated position and
temperature dependences because it is possible to check the 
equalizations after applying the corrections, it is generally
more difficult to evaluate the accuracy of the absolute calibration.
One key test to confirm the validity of the absolute calibration of the energy conversion factor
is to assess the consistency between proton and helium data. 
This is because the A/Z difference between protons and helium results in different primary energy distributions at equivalent rigidity cutoffs and also because the different signal-to-noise ratio 
will affect the event selection of penetrating particles 
should there be any such dependences. 
As shown in Fig.\ref{fig:phe}, excellent agreement was obtained between conversion
factors obtained from proton and helium MIP data. 
This figure plots the conversion factors obtained in the case of proton MIPs divided by those generated from helium MIP data
for all TASC logs. From the resulting distribution, 
it is concluded that, on average, the conversion factors agree within 0.1\%, and 
the observed deviations from unity are slightly larger than the combined errors in the conversion factors 
including the uncertainty from the energy distribution of the used events, which is studied in the following. 
To account for this small inconsistency, an additional calibration error of 1.0\% is allocated as a systematic uncertainty. 

To directly evaluate the effects of the energy distribution of incoming particles,
the MIP peak variations due to the rigidity cutoff were compared between the helium flight and MC data.  
Both data displayed similar trends, 
although there were small discrepancies at the low cutoff region, 
where low energy particles play an important role.
This could result from inaccuracy of the solar modulation parameter or
insufficient MC statistics.
Herein, a conservative estimate of a potential discrepancy of 1.0\% is introduced for the 
systematic uncertainty in the energy scale and in the calibration accuracy. 

\subsection{Linearity Measurements over the Entire Dynamic Range}
\label{uvlaser}
\begin{figure}[hbt!]
\begin{center}
\includegraphics[width=0.98 \linewidth]{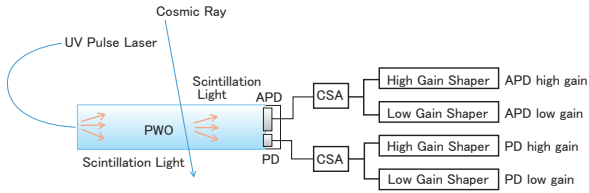}
\caption{Schematic view of UV pulse laser light injection into the PWO, together with 
TASC APD/PD readouts.}
\label{fig:apdpd}
\end{center}
\end{figure}

It was necessary to determine the input-output relationship over each gain range 
with ground-based measurements prior to launch 
because such measurements are no longer possible in orbit. 
However, the relative gain change between the four ranges can be monitored and 
corrected using gain ratio measurements, as explained in the following section.
UV pulse laser calibrations were performed on ground for linearity confirmation. 
While scanning the pulse laser intensity through six orders of magnitude, detailed measurements were made of the four APD/PD output responses
from each of the 176 PWO logs.
Figure~\ref{fig:apdpd} provides a schematic diagram of 
the UV pulse laser injection into the PWO, from the 
opposite end to the APD/PD. Since the UV photons are absorbed within a very short
distance, all the photons seen by the APD/PD are the result of PWO scintillation, which has a very
similar spectrum as that generated by charged particles.
Figure~\ref{fig:apdpd} also shows the 
hybrid APD/PD package and subsequent readout system.
By combining four readouts,
the full dynamic range of six orders of magnitude is covered while maintaining 
a nearly one order of magnitude overlap between adjacent gain ranges.
It should be noted that there is crosstalk from the APD to the PD due to stray capacitance between these two devices. 
When the charge sensitive amplifier (CSA) of the APD is saturated, 
the feedback from the CSA becomes insufficient and the potential at the APD-CSA input has a non-zero value,
which induce a signal in the PD. 
Although the crosstalk amounts to only $\sim$0.1\% of the charge ratio, it can become significant due to the APD-PD gain/area ratio of 1000 to 1 (APDs have a 20 times larger area and a 50 times higher gain). 
Since the crosstalk signal is 
proportional to the input charge and is stable, 
it is possible to calibrate the input-output relationship using UV pulse laser data. 

\begin{figure*}[hbt!]
\begin{center}
\begin{minipage}{0.56\hsize}
\includegraphics[width=0.99 \linewidth]{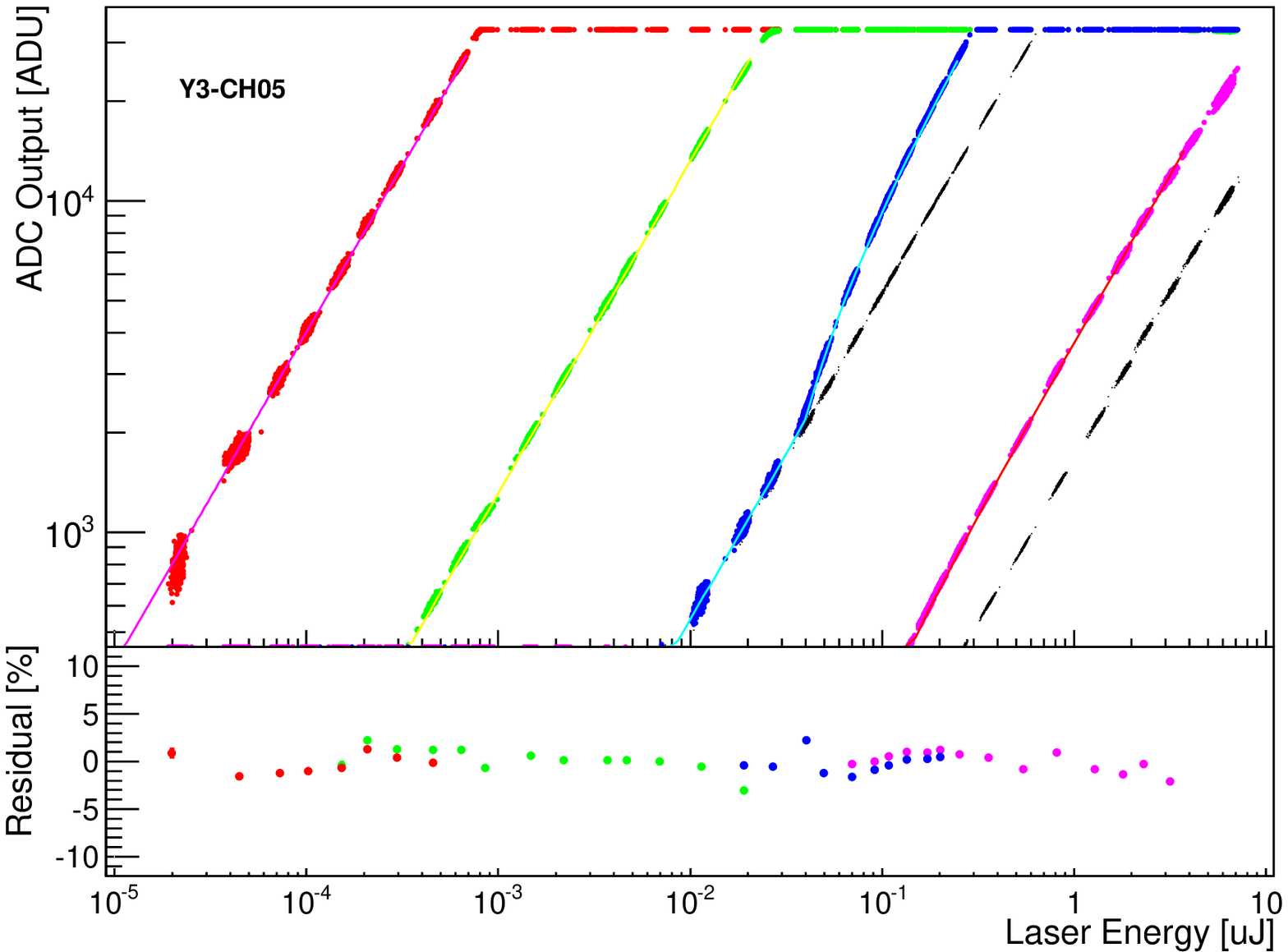} 
\end{minipage}
\begin{minipage}{0.43\hsize}
\includegraphics[width=0.99 \linewidth]{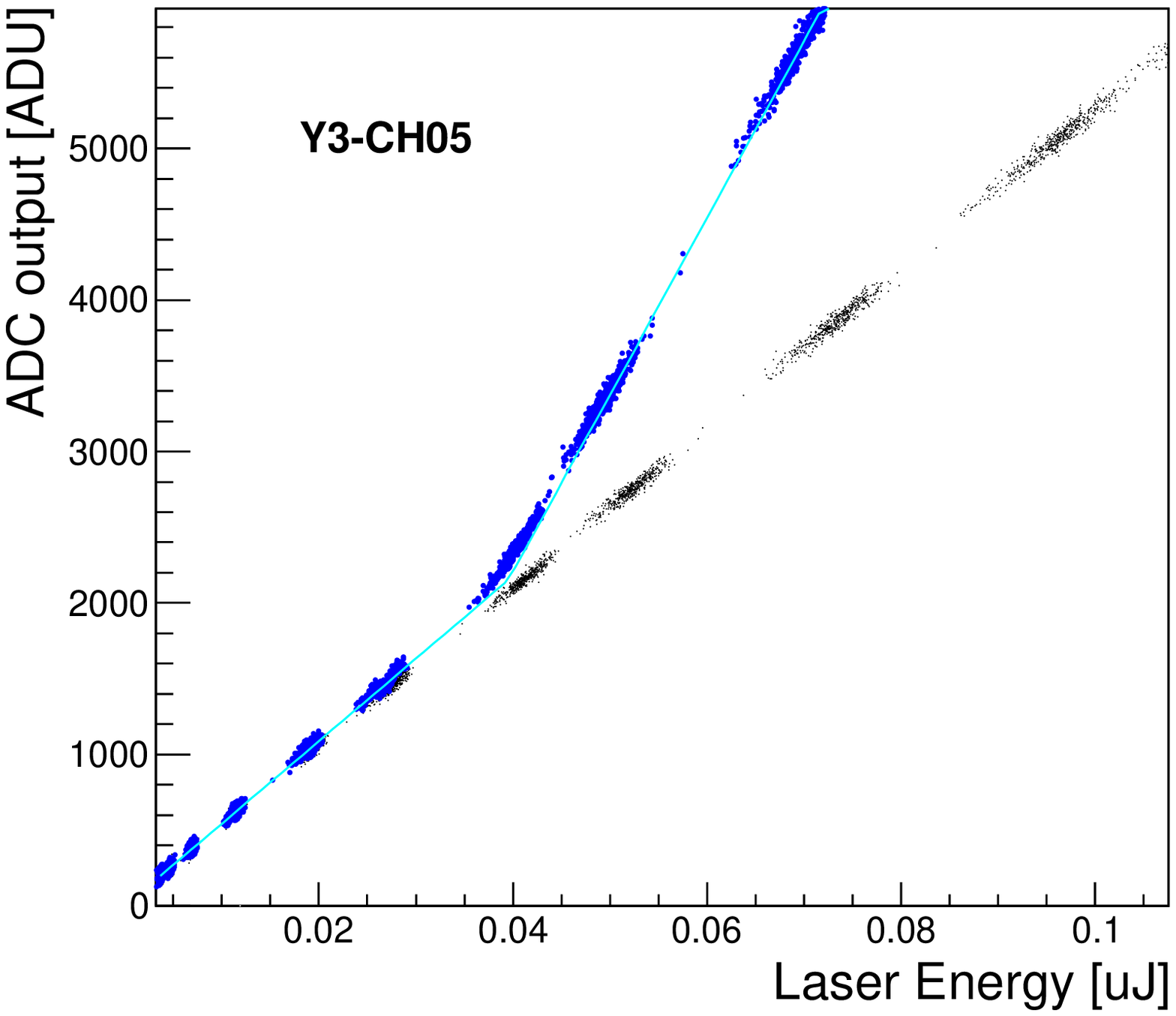} 
\end{minipage}
\caption{
Typical data acquired from UV pulse laser calibration.
({\it Left}) From left to right, 
the responses of the APD high gain (red data points), APD low gain (green), 
PD high gain (blue) and PD low gain (magenta) are plotted 
as a function of laser energy with the residuals from fitted functions shown in the 
bottom panel.
Black points in the PD range represent data acquired without an APD bias. 
({\it Right}) Close up view of the PD high gain response, applying a linear scale to better
demonstrate the simple broken linear relationship.
}
\label{fig:uvsche}
\end{center}
\end{figure*}

Figure~\ref{fig:uvsche} shows an example of the data obtained from UV pulse laser measurements. 
Here, the horizontal and vertical axes represent the laser energy and ADC values, respectively.
Since the laser energy is monitored on a pulse-by-pulse basis, the linearity over the entire dynamic range was confirmed
using 17,000 points of laser pulse data for each channel. 
As a result of the APD/PD crosstalk,
the PD response exhibits a slope break corresponding to the APD-CSA saturation point,
as shown in the {\it right} panel of Fig.~\ref{fig:uvsche}.
The responses of all APD/PD channels were measured and the data confirmed the linear 
and broken linear relationships of the APD and PD response functions, respectively,
as a result of fitting of the data points with appropriate functions for each range. 
To estimate the errors resulting from fitting the linear and broken linear functions, 
the distributions of residuals from the fitting functions were assessed for each gain range. 
From the RMS of each distribution, the errors were estimated to be 
1.4\%, 1.5\%, 2.5\% and 2.2\% for 
the APD high gain, APD low gain, PD high gain and PD low gain, respectively.
Although these errors include both possible nonlinearities and expected statistical variations in measurements, 
in addition to UV laser system calibration errors, we adopted these values as 
the actual errors due to possible non-linear effects. 

It is expected that both the APD gain and the PWO light yield will vary between on-ground conditions and
those onboard the ISS, as well as with time during on-orbit observations.
This corresponds to a change in the amount of crosstalk charge per unit energy deposit
and thus results in a slope change in the APD/PD crosstalk region.
We confirmed this effect using UV laser data acquired at a higher APD bias ($\sim \times$2 gain).
During this laser calibration process, three data sets with different APD gains 
(nominal gain, $\sim \times$2 gain and small gain without APD bias) were obtained to validate our simple model for correcting APD/PD crosstalk and to estimate the correction errors, as well as to calibrate all the gain ranges. 
This effect is revisited in the next section in relation to gain 
correlation measurements.

\subsection{Cross Calibration of Adjacent Gain Ranges}
\label{ratio}
\begin{figure*}[hbt!]
\begin{center}
\begin{minipage}{0.33\hsize}
\includegraphics[width=0.99 \linewidth]{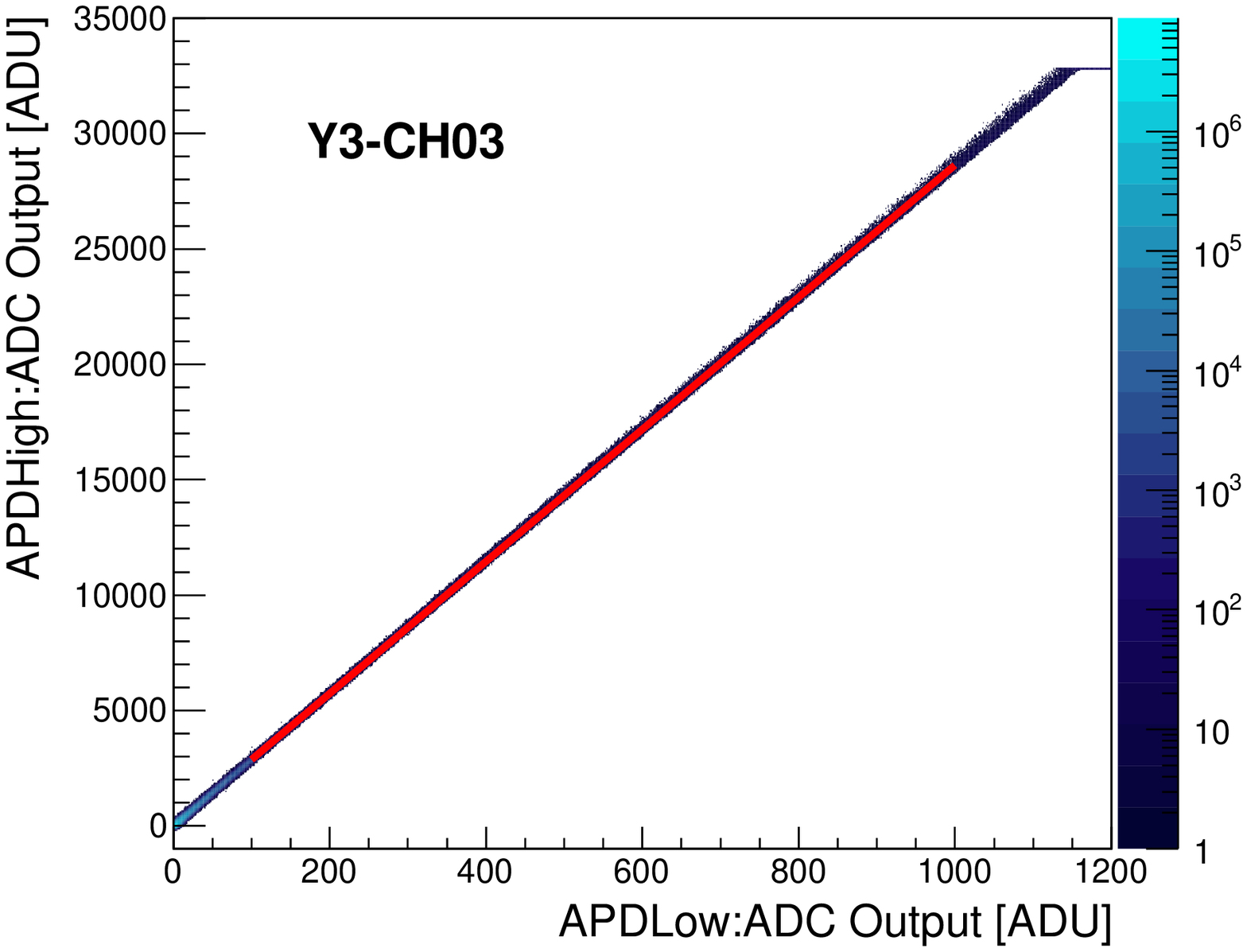}
\end{minipage}
\begin{minipage}{0.33\hsize}
\includegraphics[width=0.99 \linewidth]{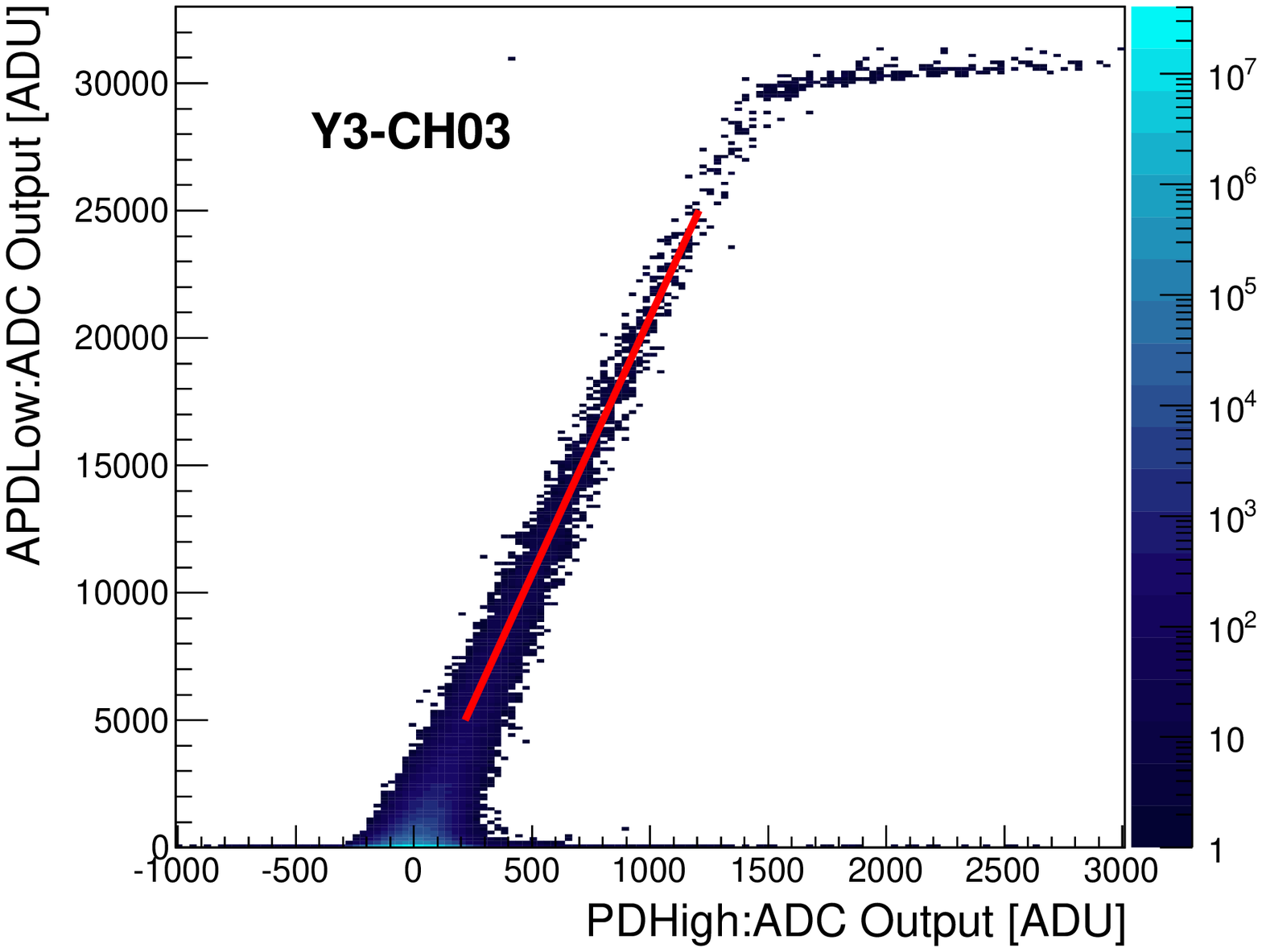}
\end{minipage}
\begin{minipage}{0.33\hsize}
\includegraphics[width=0.99 \linewidth]{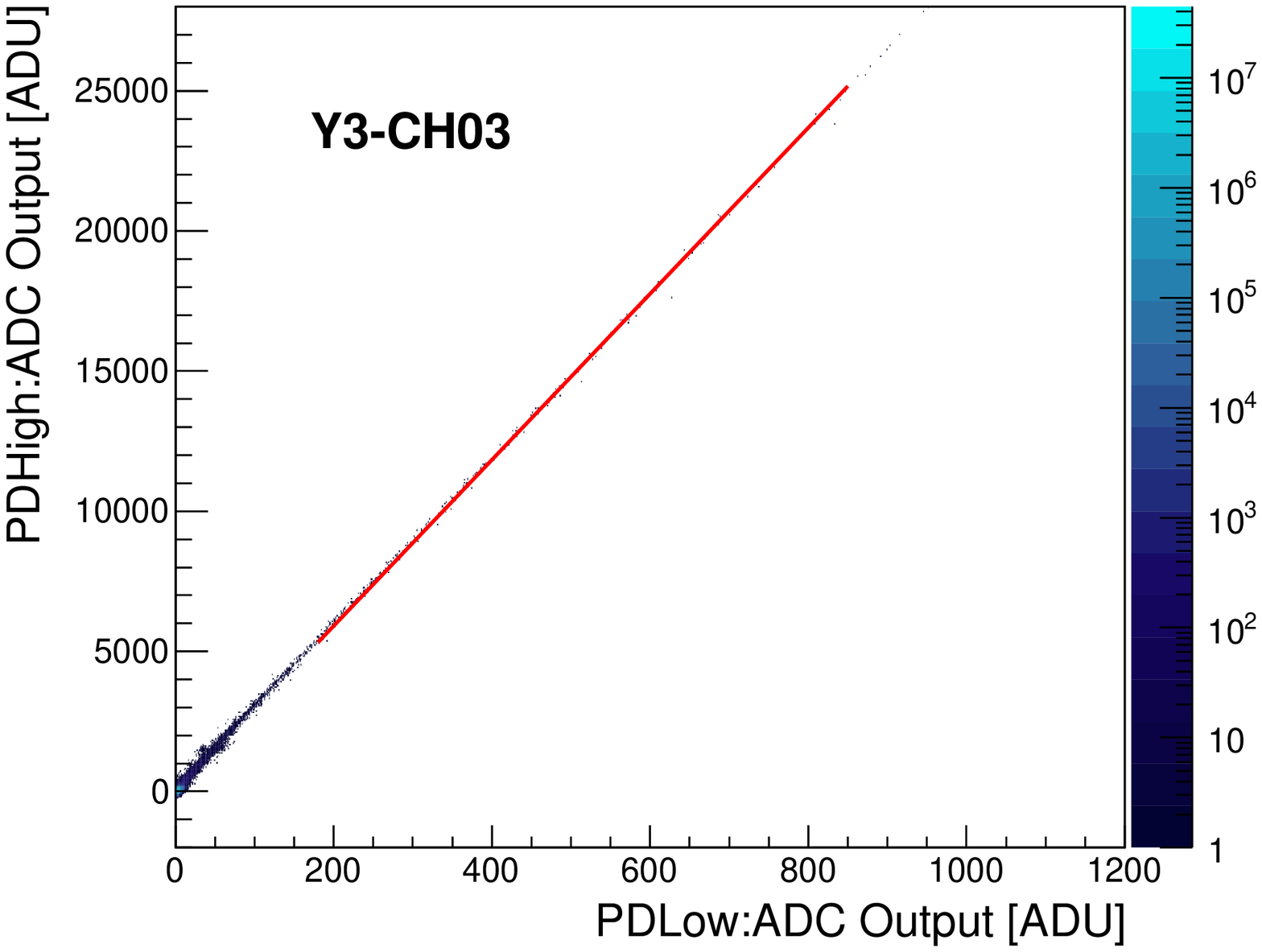}
\end{minipage}
\caption{Typical gain correlation from flight data between adjacent gain ranges. 
{\it (Left)} APD high gain to APD low gain, 
{\it (Middle)} APD low gain to PD high gain, and  
{\it (Right)} PD high gain to PD low gain regions. 
}
\label{fig:gain_ratio}
\end{center}
\end{figure*}

The gain correlations between adjacent gain ranges were used to correct for possible gain changes 
between the UV laser calibrations performed on the ground and 
observations onboard the ISS. 
Figure~\ref{fig:gain_ratio} presents examples of gain ratio measurements in
the APD high gain to APD low gain {\it (Left)}, APD low gain to PD high gain {\it (Middle)}, 
and PD high gain to PD low gain {\it (Right)} regions. 
Taking advantage of the nearly one order of magnitude overlap between adjacent gain ranges,
the same energy deposit was measured with two gain ranges and the gain ratio required to connect the two gain
ranges was determined by fitting the profile with a simple linear function.
In the fitting of each channel, proper selection of the fitting range was vital in order to avoid saturation effects in the higher gain range and the lower signal-to-noise region 
due to pedestal noise in the lower gain range. 
While the offset was set to zero in most cases, non-zero offsets during linear fitting 
were allowed in some caces involving PD-high to APD-low gain ratio fitting due to APD-to-PD 
crosstalk prior to APD-CSA saturation. 
In such cases, correct treatment was ensured by using the same offset during linear fitting of the UV pulse laser data. 
The errors on the gain ratios were determined from the parameter errors in the linear fittings 
since the reduced chi-squared distributions were found to be reasonable, having average values of approximately 1.
The errors on the ratios were found to be 0.1\%, 0.7\% and 0.1\% 
for the APD high gain to APD low gain,
PD high gain to APD low gain and PD high gain to PD low gain regions, respectively.

As explained in the previous section, slope changes 
in the on-orbit calibration of the APD range with respect to the ground data 
were foreseen due to the different environment experienced in orbit. 
This also affects the APD-to-PD crosstalk region, which was corrected based on the assumption that 
the slope change in the PD range after APD-CSA saturation 
is proportional to the slope change in the APD range between ground and 
orbit. 
When applying such corrections, it is important to identify the crosstalk component because the slope associated with the
PD gain is not affected by the APD gain change. 
UV laser data acquired with a $\sim \times$2 gain were used to validate the correction method and to estimate the associated errors.
By applying the same procedure to $\sim \times$2 gain data and comparing the 
predicted slope with the measured slope in the PD high gain range above APD-CSA saturation point, 
we were able to estimate the errors associated with our simple model for the correction of 
the APD-PD crosstalk effect.
When applying this method to on-orbit data, the error was scaled to 
the actual in-flight gain difference of $\sim$10\%, and 
the resultant error on the gain was estimated to be 1.1\%. 
Since it is not possible to determine this error from
the on-orbit data, we consider this error to represent a systematic uncertainty on the energy scale as well as 
an estimation of the calibration error that affects the energy resolution.

\begin{figure}[hbt!]
\begin{center}
\includegraphics[width=0.98 \linewidth]{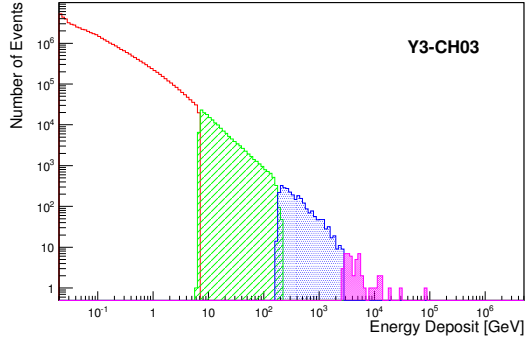}
\caption{A typical energy deposit spectrum after applying full calibration. 
Open red, hatched green, dotted blue and filled magenta 
histograms correspond to APD high, APD low, 
PD high and PD low gain ranges, respectively.
}
\label{fig:edepspec}
\end{center}
\end{figure}

Since the UV laser tests confirmed the linearity of each gain range, calibration over
the entire dynamic range is now possible by applying the conversion factor to the subsequent gain range 
using the gain ratios. 
Figure~\ref{fig:edepspec} shows a typical calibrated energy deposit spectrum for one
TASC channel. 
A smooth transition between adjacent gain ranges is clearly observed. 
To determine the errors due to extrapolation from the region in which the gain ratio was 
measured to the uppermost point in each gain range, the slopes were compared between
the gain ratio region and the full range for each gain range by employing UV laser data. 
Using the RMS of the distribution of the relative slope changes
obtained from all the TASC channels, 
the errors were 
estimated as 1.6\% and 1.8\% for the APD high gain to APD low gain and PD high gain 
to PD low gain regions, respectively. 
Note that the RMS is dominated by the 
UV laser test statistics which is limited especially in the 
overlapping region due to shorter lever arm.
Since there is no systematic shift in their slopes, these extrapolation errors can be considered
as a part of the calibration accuracy, rather than a component of 
the energy scale uncertainty. 
The APD low gain to 
PD high gain extrapolation error was estimated at a higher value 2.0\%, 
to account for possible gain changes relative to the on-ground calibration. 
This conservative error estimate should be considered as a component of the energy scale uncertainty. 

\section{Energy Measurement: Error and Resolution}
\label{disc}

Table~\ref{tab:error_budget} summarizes the error budget for CALET energy measurements
based on the discussions in the previous sections.
Note that the systematic error in the energy measurements resulting from the 
MC simulation based on EPICS is negligible below the energy of
95\% containment for electromagnetic showers ($\sim$20~TeV). 
Highly detailed detector geometries and materials were employed in our MC simulation, based on 
the CAD model for the CALET detector. 
\begin{table}[htb!]
\begin{center}
 \caption{Summary of the error budget in the energy calibration.} 
 \label{tab:error_budget}
 \begin{tabular}{ccllc} \hline
MIP &  & Energy conversion & 2.6\% \\ \hline 
    \multicolumn{3}{l}{\hspace{0.5cm} Peak fitting of MC and flight data} & 0.6\% \\ 
    \multicolumn{3}{l}{\hspace{0.5cm} Fitting range dependence} & 0.6\%$^{(*)}$ \\ 
    \multicolumn{3}{l}{\hspace{0.5cm} Position dependence} &  1.8\% \\ 
    \multicolumn{3}{l}{\hspace{0.5cm} Temperature dependence} & 1.0\% \\ 
    \multicolumn{3}{l}{\hspace{0.5cm} Rigidity cutoff dependence} & 1.0\%$^{(*)}$ \\ 
    \multicolumn{3}{l}{\hspace{0.5cm} Systematic uncertainty estimated} & \\
    \multicolumn{3}{l}{\hspace{1.2cm} from p/He consistency} & 1.0\% \\ \hline 
UV Laser &  & Linearity & 1.4$\sim$2.5\% \\ \hline 
    \multicolumn{3}{l}{\hspace{0.5cm} Fit error} & \\
    \multicolumn{3}{l}{\hspace{1cm} APD high gain} & 1.4\% \\ 
    \multicolumn{3}{l}{\hspace{1cm} APD low gain} & 1.5\% \\ 
    \multicolumn{3}{l}{\hspace{1cm} PD high gain} & 2.5\% \\ 
    \multicolumn{3}{l}{\hspace{1cm} PD low gain} & 2.2\% \\ \hline 
Gain Ratio &  & Gain range connection & 1.6$\sim$2.1\% \\ \hline 
     \multicolumn{3}{l}{\hspace{0.5cm} Fit error} \\
     \multicolumn{3}{l}{\hspace{1cm} APD-high to APD-low gain} & 0.1\% \\
     \multicolumn{3}{l}{\hspace{1cm} APD-low to PD-high gain} & 0.7\% \\ 
     \multicolumn{3}{l}{\hspace{1cm} PD high to PD low gain} & 0.1\% \\ 
     \multicolumn{3}{l}{\hspace{0.5cm} Slope extrapolation} \\
     \multicolumn{3}{l}{\hspace{1cm} APD-high to APD-low gain} & 1.6\% \\ 
     \multicolumn{3}{l}{\hspace{1cm} APD-low to PD-high gain} & 2.0\%$^{(*)}$ \\ 
     \multicolumn{3}{l}{\hspace{1cm} PD high to PD low gain} & 1.8\% \\ \hline 
\multicolumn{3}{l}{Sampling Bias}  & 0.5\%$^{(**)}$ \\ \hline 
    \multicolumn{4}{l}{\hspace{0.01cm} $^{(*)}$ also considered as systematic error on energy scale}\\
    \multicolumn{4}{l}{\hspace{0.01cm} $^{(**)}$ energy-scale systematic error only}\\
 \end{tabular}
\end{center}
\end{table}

\begin{figure}[hbt!]
\begin{center}
\includegraphics[width=0.99 \linewidth]{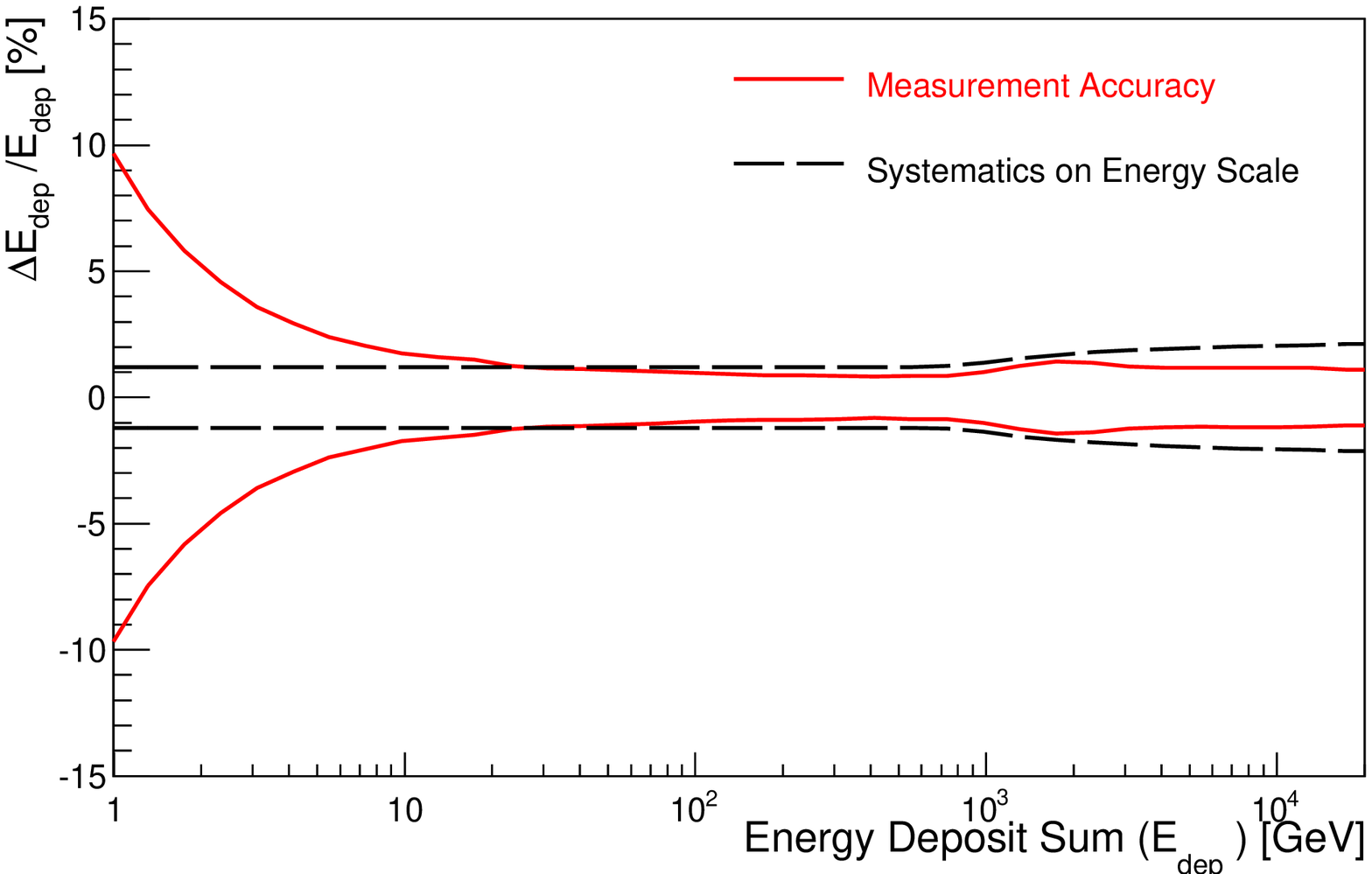}
\includegraphics[width=0.99 \linewidth]{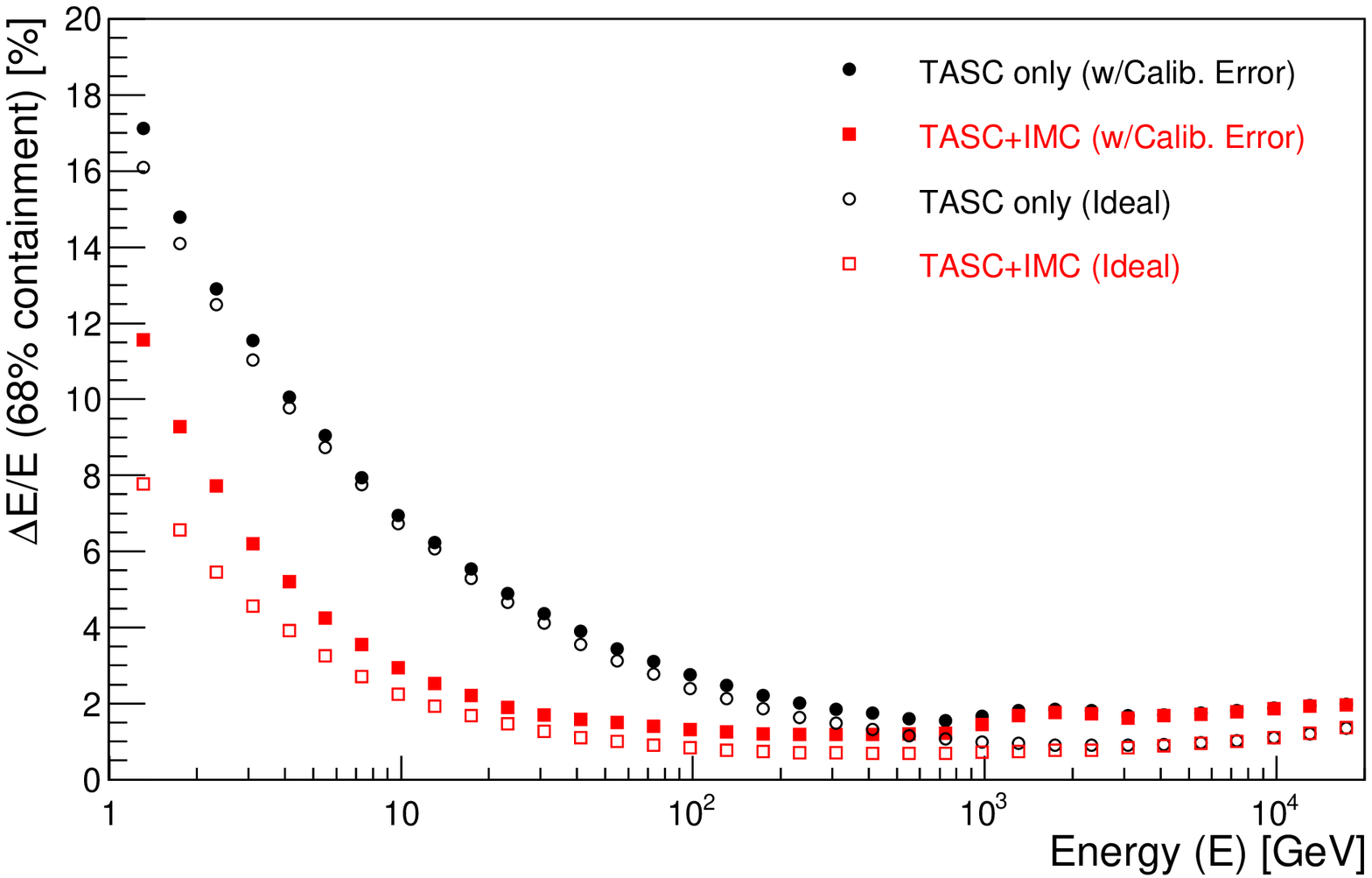}
\caption{
(Top) Energy dependence of the relative error in the energy deposit sum measurements
for electrons, considering all the energy calibration errors and detector responses (solid red lines). The systematic uncertainty on an absolute scale is also shown by black dashed lines. (Bottom) Estimated energy resolution for electrons as a function of energy. 
Open red squares denote intrinsic resolution and closed red squares denote actual resolution
including all the detector responses and calibration errors in the case of 
energy determination using the TASC$+$IMC.
Circular symbols indicate energy determination using the TASC only.
} 
\label{fig:syst}
\end{center}
\end{figure}

Using the estimated calibration errors and measured detector responses,
such as the pedestal noise on a channel-by-channel basis,
the errors in the energy deposit sum were calculated for simulated
electron events from 1~GeV to 20~TeV.
The top panel of Fig.~\ref{fig:syst} presents the energy dependence of the relative error in the energy deposit sum measurements. 
As clearly shown by this figure, a 2\% precision level energy calibration was achieved 
over the entire dynamic range above 10~GeV. The reduced accuracy with which the energy deposit
can be determined below 10~GeV is due to pedestal noise. 
As reported in detail in Ref.~\cite{Niita15}, 
the requirements for the calibration error of each TASC log can be relaxed by a factor of $\sim$3
compared to that for the energy resolution, as long as these individual errors 
of in total $\sim$6\% are randomly distributed. 
This is due to the fact that, on average, $\sim$10 TASC logs contribute significantly 
to an event's energy measurement. 
The results obtained here are therefore perfectly consistent with the expected values. 

The estimated systematic uncertainty is also plotted on an absolute scale in
Fig.~\ref{fig:syst}.
The systematic uncertainty in the energy scale was estimated 
to be less than $\sim$2\%. 
Since the calibration error is a fixed value for each channel, there could be
systematic bias on the energy measurements.
To account for this effect, several sets of simulation data were generated 
and evaluated for such a systematic bias
by calculating the ratio from estimated energy 
deposit sum to true energy sum.
The resultant error was estimated in an energy dependent manner and 
found to be $\le$0.5\% as indicated as 'Sampling Bias' in Table\ref{tab:error_budget}.
It should be noted that the
PD range becomes important, i.e., accounts for more than 20\%  
of an energy measurement, 
at an energy deposit sum of 1~TeV, 
resulting in slightly larger systematic uncertainties in this range,
although the calibration accuracy is still satisfactory.
Furthermore,
improvement in our knowledge of the systematic uncertainty on the energy scale 
is expected as long as the collected data statistics grows, which will allow us 
to understand the detector better. 

To conclude, 
the estimated energy resolution for electrons as a function of energy is plotted
in the bottom panel of Fig.~\ref{fig:syst}. 
Thanks to the detailed calibration process described in this paper, 
a very high energy resolution has been achieved over the entire dynamic range. 

\section{Conclusion}
\label{conc}
Energy calibration of the CALET, 
launched to the ISS in August 2015 and
accumulating scientific data since October 2015, 
was performed using both flight data and calibration 
data acquired on the ground before launch.
By taking advantage of the fully-active total 
absorption calorimeter, absolute calibration between ADC
units and energy was possible with an accuracy of a few percent, using penetrating particles.
Successful calibration was achieved over the complete dynamic range of six orders 
of magnitude for each TASC channel
with sufficient accuracy to maintain a fine resolution of 2\% above 100~GeV 
by combining two calibration processes:
linearity measurements over each gain range and 
determination of the correlation between adjacent gain ranges.
The systematic error in the energy scale was also estimated based on the calibration results and
was found to be $\le$ 2\%. 
Based on long duration observations of high energy cosmic rays onboard the ISS,
the measurement of the inclusive ($e^{+} + e^{-}$) electron spectrum 
well into the TeV region with unprecedented accuracy is expected,
as well as measurements of gamma-rays, protons and nuclei. 





\section*{Acknowledgements}
We gratefully acknowledge JAXA's contributions to the development of CALET 
and to operations on board the ISS. We also wish to express our sincere gratitude to ASI and NASA for their support of the CALET project. 
Finally, this work was partially supported by a JSPS Grant-in-Aid for Scientific Research (S) (no. 26220708) and by the MEXT-Supported Program for the Strategic Research Foundation 
at Private Universities (2011-2015) (no. S1101021) at Waseda University.

\end{document}